\documentclass[12pt,reqno]{amsart}

\usepackage{amsmath,amsfonts,amsthm,amssymb}

\usepackage[height=20cm,top=24mm,bottom=24mm,left=26mm,right=26mm]{geometry}

\usepackage[numbers,sort]{natbib}
\usepackage{hyperref}
\usepackage[]{graphicx}                                                    
\numberwithin{equation}{section}
\newcommand{\I}{\mathrm{i}}
\newcommand{\E}{\mathrm{e}}

\DeclareMathOperator{\sign}{sgn}
\DeclareMathOperator{\Tr}{Tr}

\DeclareMathOperator{\ch}{ch}
\DeclareMathOperator{\erfc}{erfc}

\DeclareMathDelimiter{\Norm}{\mathord}{largesymbols}{"3E}{largesymbols}{"3E}

\DeclareMathOperator{\Coeff}{Coeff}
\def\clap#1{\hbox to 0pt{\hss#1\hss}}

\allowdisplaybreaks[4]
\begin{document}
\baselineskip 16pt
\parskip 8pt
\sloppy

%%%%%%%%%%%%%%%%%% TITLE %%%%%%%%%%%%%%

 \title[]{$\mathcal{N}=4$ Superconformal Algebra and the Entropy of
   HyperK{\"a}hler Manifolds}

%%%%%%%%%%%%%%%%%%%%%%% AUTHOR(S) %%%%%%%%%%%%%%%%%%%

\author[T. Eguchi]{Tohru \textsc{Eguchi}}

\author[K. Hikami]{Kazuhiro \textsc{Hikami}}

%%%%%%%%%%%%%%%%%%%%%%% ADDRESS %%%%%%%%%%%%%%%%%%%%

\address{Yukawa Institute for Theoretical Physics, Kyoto University,
  Kyoto 606--8502, Japan}
\email{
  \texttt{eguchi@yukawa.kyoto-u.ac.jp}
}

\address{Department of Mathematics, 
  Naruto University of Education,
  Tokushima 772-8502, Japan.}

\email{
%  \texttt{hikami@naruto-u.ac.jp}
  \texttt{KHikami@gmail.com}
}

%%%%%%%%%%%%%%%%%%%%%% DATE %%%%%%%%%%%%%%%%%%%%%%%%%%%
%(Received: \hspace{40mm})

%\vspace{18pt}
\date{\today}
%\date{September 2, 2009.}

%%%%%%%%%%%%%%%%%%%%%% ABSTRACT %%%%%%%%%%%%%%%%%%%%%%
\begin{abstract}
We study the elliptic genera of hyperK{\"a}hler manifolds using the
representation theory of $\mathcal{N}=4$ superconformal algebra.
We consider the decomposition of the elliptic genera in terms of
$\mathcal{N}=4$ irreducible characters, and derive the rate of
increase of the multiplicities of half-BPS representations making use
of  
Rademacher expansion. Exponential increase of the multiplicity
suggests that we can associate the notion of an entropy to the
geometry of hyperK\"ahler manifolds. In the case of symmetric products
of $K3$ surfaces our entropy agrees with the black hole entropy of
D5-D1 system.
\end{abstract}

%%%%%%%%%%%%%%%%%%%%%%% Key Words %%%%%%%%%%%%%%%%%%%%%%%%%%

\keywords{
}

\subjclass[2000]{
}

%%%%%%%%%%%%%%%%%%%%%%%%%%%%%%%%%%%%%%%%%%%%%%%%%%%%%%%%%
%\newpage

%%\renewcommand{\thefootnote}{\arabic{footnote}}
\maketitle
%%%%%%%%%%%%%%%%%%%%%%%%%%%%%%%%%%%%
\section{Introduction}

It has been known for some time ~\cite{EgucTaor88b}
that characters of the BPS representations of the extended superconformal algebra 
do not in general have a good modular property.
This is   because of the existence of special singular vectors coming
from the BPS condition ($G^i|h\rangle=0$). 
Thus BPS characters are not
regular theta functions but are mock (pseudo) theta functions of the
kind first introduced by Ramanujan~\cite{GEAndre89a,Dyson88walk}.
Systematic understanding of
mock theta functions, however,  was not available until very recently.
Intrinsic structure behind them was first revealed by
Zwegers several years ago ~\cite{Zweg02Thesis},
and they are identified as the holomorphic part of the harmonic
Maass forms
(see Appendix for definition).
Since the work of Zwegers,
the theory of the mock theta function
has been applied to the
theory
of partitions~\cite{BrinKOno06a},
and its relationship with the quantum invariant for links
and
3-manifolds has been clarified ~\cite{LawrZagi99a,KHikami03c,KHikami04b,KHikami05b,KHikami05a,KHikami06b,DZagie01a}
(see Ref.~\citenum{KOno08a} for a review on recent developments).

This paper is a sequel to our previous 
papers~\cite{EguchiHikami08a,EguchiHikami09a},
where we have studied representation theory of the $\mathcal{N}=4$ superconformal algebras using the method of Zwegers  and  in particular the decomposition of the elliptic genus of the $K3$ surface 
in terms of irreducible characters of $\mathcal{N}=4$ algebra. 

In general the  elliptic genus of hyperK\"ahler manifold of
complex-dimension $2k$ has an expansion
\begin{multline}
  \mbox{elliptic genus}
  =
  \sum_{\ell \in
    \left\{ 0, \frac{1}{2}, \dots, {k\over 2} \right\}
  }
  c_{\ell} \, 
  \left[
    \mbox{BPS representation}:h={k\over 4},\ell
  \right]
  \\
  +
  \sum_{n=1}^{\infty}
%  \sum_{\ell={1\over 2}}^{k\over 2}
  \sum_{\ell
    \in \left\{
      {1\over 2},1,\dots, {k\over 2}
    \right\}
  }
  p_k^{(\ell)}(n) \,
  \left[
    \mbox{non-BPS representation}: h=n+{k\over 4},\ell
  \right].
  \label{elliptic-genus-decompose}
\end{multline}
Here $h$ and $\ell$ respectively denote the conformal dimension and
isospin of highest weight states.
In this paper we introduce the Rademacher
expansion and determine the asymptotic behavior of the multiplicity
factors $p_k^{(\ell)}(n)$ as $n$ becomes large.
We shall show that they have an exponential growth and at  large $k$
behave as 
\begin{equation}
  \label{asymptotic_p_intro}
  p_k^{(\ell)}(n)
%  \approx 
  \sim
  \exp
  \left(
    2 \, \pi \, \sqrt{k \, n-\ell^2}
  \right) .
\end{equation}
Such an exponential behavior of the degeneracy is reminiscent of the
entropy of black holes.

In the elliptic genus the right-moving sector is held fixed at the Ramond 
ground state and hence the non-BPS states
in~\eqref{elliptic-genus-decompose} are actually the half-BPS states
(BPS (non-BPS) in the right-(left-)moving sector).
Counting the
asymptotic degeneracy of
states protected by supersymmetry
amounts to computing the
entropy of  systems. 
Actually as we see below, when one considers the case of symmetric
product of $K3$ surfaces 
$K3^{[k]}$  it in fact  agrees with the entropy of the standard D5-D1
black holes in
$AdS^3\times S^3\times K3$~\cite{StroVafa96a,CvetLars98a}. 
Positivity inside the square root of~\eqref{asymptotic_p_intro}
corresponds to the cosmic censorship in classical general
relativity~\cite{CvetLars98a,DijMalMooVer00a}.

We propose in this paper that arbitrary hyperK\"ahler manifolds carry
entropy as defined above. In~\eqref{elliptic-genus-decompose} a bad
modular property of BPS characters is exactly compensated by the equally bad
modular property of the infinite series
$\sum_n p_k^{(\ell)}(n) \, q^n$. 
Thus the lack of modular behavior of  BPS
characters is the origin of entropy in hyperK\"ahler manifolds. 

This paper is organized as follows.
In Section~\ref{sec:review}
we briefly review our previous
results in Refs.~\citenum{EguchiHikami08a,EguchiHikami09a}.
In Section~\ref{sec:Poincare} we study the Rademacher expansion of the
Fourier coefficients of the
vector-valued harmonic Maass form by use of the Poincar{\'e}--Maass
series.
In Section~\ref{sec:decomposition} we  study  the 
decomposition of the elliptic genera of the hyperK{\"a}hler
manifolds in terms of irreducible characters.
By use of the Rademacher expansion, we derive the asymptotic behavior of
the multiplicity of the non-BPS representations.
We present the cases of level-$2$ and -$3$ in some detail.
The last section contains concluding remarks.

%%%%
\section{Superconformal Algebras and Mock Theta Functions}
\label{sec:review}
\subsection{Characters of Superconformal Algebras}
The $\mathcal{N}=4$ superconformal algebra at level $k$ has a central
charge
$c=6 \, k$, and contains an affine $SU(2)_k$ algebra.
Its highest
weight state
$|\Omega\rangle$ is labeled by the
conformal weight $h$ and the isospin $\ell$,
\begin{equation*}
  \begin{gathered}
    L_0 \, | \Omega \rangle = h \, | \Omega \rangle ,
    \\
    T_0^3 \, | \Omega \rangle = \ell \, | \Omega \rangle.
  \end{gathered}
\end{equation*}
The character of a representation is defined by 
\begin{equation}
  \ch_{k,h,\ell} ( z; \tau)
  =
  \Tr_{\mathcal{H}}
  \left(
    \E^{4  \, \pi \,  \I \,  z \,  T_0^3} \,
    q^{L_0 - \frac{c}{24}}
  \right) ,
\end{equation}
where $q=\E^{2 \, \pi \, \I \, \tau}$ with $\tau\in\mathbb{H}$, and
$\mathcal{H}$ denotes the Hilbert space of the representation.
In the following
we often use $\zeta =\E^{2 \, \pi \, \I \, z}$
with $z \in \mathbb{C}$. In 
$\mathcal{N}=4$ theory
we have two types of
representations~\cite{EgucTaor86a,EgucTaor88a,EgucTaor88b};
massless (BPS) and massive (non-BPS) representations. 
In the Ramond sector, their character formulas are given
as follows;
\begin{itemize}
\item massless representations
  ($h=\frac{k}{4}$, and $\ell=0,\frac{1}{2}, \dots, \frac{k}{2}$),
  \begin{equation}
    \ch^R_{k,\frac{k}{4},\ell}(z;\tau)
    =
    \frac{\I}{\theta_{11}(2 z; \tau)} \cdot
    \frac{
      \left[ \theta_{10}(z;\tau) \right]^2
    }{
      \left[ \eta(\tau) \right]^3
    }
    \sum_{\varepsilon=\pm 1} \sum_{m \in \mathbb{Z}}
    \varepsilon \,
    \frac{
      \E^{4 \pi \I \varepsilon \left( (k+1)m+\ell \right) z}
    }{
      \left(
        1+ \E^{- 2 \pi \I \varepsilon z} \, q^{-m}
      \right)^2
    } \,
    q^{(k+1) \, m^2 + 2 \, \ell \, m} 
    ,
    \label{define_massless_ch}
  \end{equation}

\item massive representations
  ($h > \frac{k}{4}$ and $\ell=\frac{1}{2}, 1, \dots, \frac{k}{2}$),
  \begin{equation}
    \label{massive_character}
    \ch^R_{k,h,\ell}(z;\tau)
    =
    q^{h - \frac{\ell^2}{k+1} - \frac{k}{4}} \,
    \frac{
      \left[ \theta_{10}(z;\tau) \right]^2
    }{
      \left[ \eta(\tau) \right]^3
    } \,
    \chi_{k-1, \ell - \frac{1}{2}}(z;\tau) ,
  \end{equation}
  where $\chi_{k,\ell}(z;\tau)$ denotes the affine SU($2$) character
  \begin{equation}
    \label{define_affine_SU2}
    \chi_{k,\ell}(z;\tau)
    =
    \frac{
      \vartheta_{k+2,2\ell+1} - \vartheta_{k+2,-2\ell-1}
    }{
      \vartheta_{2,1} - \vartheta_{2,-1}
    }(z;\tau) ,
  \end{equation}
  with the theta series defined by
  \begin{align}
    \vartheta_{P,a}(z;\tau)
    & =
    \sum_{n \in \mathbb{Z}} q^{\frac{(2 P n +a)^2}{4 P}} \,
    \E^{2 \pi \I z (2 P n +a)} .
  \end{align}
  Note that the denominator of the affine character equals 
  \begin{equation}
    \left( \vartheta_{2,1} - \vartheta_{2,-1} \right)(z;\tau)
    =
    - \I \, \theta_{11}(2 \, z; \tau) .
  \end{equation}

\end{itemize}
Characters in other sectors are obtained by spectral flow:
$z\rightarrow z+{1\over 2} \,(\widetilde{R})$,
$z\rightarrow z+{\tau\over 2} \,(NS)$,
$z \rightarrow z+{1+\tau\over 2} \, (\widetilde{NS})$. 
  
At the unitarity boundary $h=\frac{k}{4}$, the non-BPS representation
decomposes into a sum of the BPS representations.
For instance, in
the $\widetilde{R}$ sector ($R$ sector with $(-1)^F$
insertion) we have 
\begin{equation}
  \begin{aligned}[b]
    \lim_{h \searrow \frac{k}{4}} \ch^{\widetilde{R}}_{k,h,\ell}(z;\tau)
    & =
    (-1)^{2\ell+1} \,
    q^{- \frac{\ell^2}{k+1}} \,
    \frac{
      \left[ \theta_{11}(z;\tau) \right]^2
    }{
      \left[ \eta(\tau) \right]^3
    } \,
    \chi_{k-1,\ell - \frac{1}{2}}(z;\tau)
    \\
    & =
    \ch^{\widetilde{R}}_{k,\frac{k}{4},\ell}(z;\tau)
    +
    2 \,
    \ch^{\widetilde{R}}_{k,\frac{k}{4},\ell-\frac{1}{2}}(z;\tau)
    +
    \ch^{\widetilde{R}}_{k,\frac{k}{4},\ell-1}(z;\tau) .
  \end{aligned}
  \label{recursion_massless_R_tilde}
\end{equation}

\subsection{Conformal Characters and Mock Theta Functions}
For notational convenience we set the holomorphic function
$C_k(z;\tau)$ to be the massless superconformal character with
isospin-$0$ in $\widetilde{R}$ sector;~\footnote{In this
  article
  we slightly
  modify the notations from our previous
  papers~\cite{EguchiHikami08a,EguchiHikami09a}.}
\begin{align}
  C_k(z;\tau)
  & =
  \ch_{k,h=\frac{k}{4},\ell=0}^{\widetilde{R}}(z;\tau) 
  \nonumber
  \\
  & =
  \frac{
    \left[\theta_{11}(z;\tau) \right]^2}{
    \left[ \eta(\tau) \right]^3
  }
  \,
  \frac{
    \I}{
    \theta_{11}(2 \, z; \tau)
  }
  \,
  \sum_{n \in \mathbb{Z}}
  q^{(k+1) \, n^2} \,
  \E^{4  \pi  \I   (k+1)   n  z} \,
  \frac{
    1+ q^n \, \E^{2 \pi \I z}
  }{
    1- q^n \, \E^{2 \pi \I z}
  } .
  \label{define_C_k}
\end{align}
Note that massless representation carries the Witten index 
\begin{equation}
  C_k(z=0;\tau)=1.
  \label{Witten_index}
\end{equation}
 It is known that the function $C_k(z;\tau)$ does not have a good
 behavior under modular transformation: one has to find its
 suitable ``completion'' which has a good modular behavior.  
The following completion of
$C_k(z;\tau)$ 
has been obtained in our previous work ~\cite{EguchiHikami08a}
\begin{equation}
  \label{general_Maass_pre}
  \widehat{C}_k(z;\tau)
  =
  C_k(z;\tau)
  -
  \frac{1}{\I \sqrt{2 \, (k+1)}} \,
  \sum_{a=1}^{k}
  R_k^{(a)}(\tau)
  \, B_k^{(a)}(z;\tau) .
%   =
%   C_P(z;\tau)
%   -
%   \frac{1}{\I \sqrt{2 \, P}}
%   \,
%   \left[
%     \frac{\theta_{11}(z;\tau)}{
%       \eta(\tau)}
%   \right]^2
%   \sum_{a=1}^{P-1}
%   \frac{
%     R_P^{(a)}(\tau)}{
%     \eta(\tau)
%   } \,
%   \frac{
%     \vartheta_{P,a}
%     -
%     \vartheta_{P,-a}
%   }{
%     \vartheta_{2,1}
%     -
%     \vartheta_{2,-1}
%   }(z; \tau) .
\end{equation}
Here the basis functions 
$B_k^{(a)}(z;\tau)$ are proportional to the massive characters of
$\mathcal{N}=4$ algebra~\eqref{massive_character}
\begin{align}
  B_k^{(a)}(z;\tau)
  &  =
  \frac{
    \left[ \theta_{11}(z;\tau) \right]^2
  }{
    \left[
      \eta(\tau)
    \right]^3
  } \,
  \chi_{k-1,\frac{a-1}{2}}(z;\tau)
  \nonumber \\
  &
  =\frac{
    \left[ \theta_{11}(z;\tau) \right]^2
  }{
    \left[
      \eta(\tau)
    \right]^3
  } \cdot
  \frac{
    \vartheta_{k+1,a} - \vartheta_{k+1,-a}
  }{
    \vartheta_{2,1} - \vartheta_{2,-1}
  }(z; \tau)  .
  \label{basis-function}
\end{align}
The non-holomorphic function $R_k^{(a)}(\tau)$ is defined as
\begin{equation}
  \label{R_error_function}
  R_k^{(a)}(\tau)
  =
  \I \sqrt{2 \, (k+1)} 
  \sum_{%\mathclap{
      \substack{
        n \in \mathbb{Z}
        \\
        n = a \mod 2(k+1)
      }}%}
  \left[
    \sign \left( n+\frac{1}{2} \right)
    -
    E \left(
      n \, \sqrt{\frac{\Im \tau}{k+1}}
    \right)
  \right] \,
  q^{-\frac{n^2}{2 (k+1)}} ,
\end{equation}
where $E(x)$ is the error function
\begin{equation*}
  E(x)
  = 2 \int_0^x \E^{- \pi \, t^2} \,
  \mathrm{d} t
  =
  1 - \erfc\left( \sqrt{\pi} \, x \right) .
\end{equation*}
The function $R_k^{(a)}(\tau)$
can be rewritten as a period integral,
\begin{gather}
  \label{define_R_k_a}
  R_k^{(a)}(\tau)
  =
  \int_{- \overline{\tau}}^{\I \infty}
  \frac{
    \Psi_k^{(a)}(z)
  }{
    \sqrt{\frac{z+\tau}{\I}}
  } \,
  \mathrm{d} z ,
\end{gather}
where $\Psi_k^{(a)}(\tau)$ denotes a vector-valued
modular form with
weight-$3/2$ proportional to the affine SU(2) character;
\begin{equation}
  \Psi_k^{(a)}(\tau) = \left[ \eta(\tau) \right]^3
  \chi_{k-1,\frac{a-1}{2}}(0;\tau)
  =  \left[ \eta(\tau) \right]^3
  \frac{
    \vartheta_{k+1,a} - \vartheta_{k+1,-a}
  }{
    \vartheta_{2,1} - \vartheta_{2,-1}
  }(0; \tau)  .
  \label{def_Psi}
\end{equation}
In the sense of Zagier~\cite{Zagier08a}, the massless superconformal
character $C_k(z;\tau)$ is a mock theta function whose shadow is
$\Psi_k^{(a)}(\tau)$.
The completion $\widehat{C}_k(z;\tau)$ is a real analytic Jacobi form
with weight-$0$ and index-$k$.
Its modular properties are summarized as follows;
\begin{equation}
  \begin{gathered}
    \widehat{C}_k(z;\tau)
    =
    \E^{-2  \pi   \I k  \frac{z^2}{\tau}} \,
    \widehat{C}_k\left(\frac{z}{\tau};-\frac{1}{\tau}\right) ,
    \\[2mm]
    \widehat{C}_k(z;\tau+1)
    =
%     \widehat{C}_P(z;\tau)
%     \\[2mm]
% %
    \widehat{C}_k(z+1 ;\tau)
    =
    \widehat{C}_k(z ;\tau) ,
    \\[2mm]
    \widehat{C}_k(z+\tau ;\tau)
    =
    q^{-k} \, \E^{-4  \pi  \I  k  z} \,
    \widehat{C}_k(z ;\tau) .
  \end{gathered}
\end{equation}
We notice that the basis function $B_k^{(a)}(z;\tau)$ 
is a vector-valued
Jacobi form with weight-($-1/2$) and index-$k$;
\begin{equation}
  \label{ST_transform_B_k}
  \begin{gathered}
    B_k^{(a)}(z;\tau)
    = - \sqrt{\frac{\tau}{\I}} \,
    \E^{- 2  \pi  \I  k  \frac{z^2}{\tau}} \,
    \sum_{b=1}^{k}
    \sqrt{\frac{2}{k+1}} \,
    \sin \left( \frac{a \, b }{k+1} \, \pi \right) \,
    B_k^{(b)} \left( \frac{z}{\tau} ; - \frac{1}{\tau} \right),
    \\[2mm]
    B_k^{(a)}(z;\tau+1)
    =
    \E^{ \frac{a^2}{2(k+1)} \pi \I} \,
    B_k^{(a)}(z; \tau),
    \\[2mm]
    B_k^{(a)}(z+1; \tau) = B_k^{(a)}(z; \tau) ,
    \\[2mm]
    B_k^{(a)}(z+\tau ; \tau)
    =
    q^{-k} \, \E^{-4  \pi  \I  k  z} \,
    B_k^{(a)}(z;\tau).
  \end{gathered}
\end{equation}

%%%%%
\subsection{Harmonic Maass Form}
Next we define the elements of a $k\times k$ matrix
$\mathbf{B}_k(\boldsymbol{z};\tau)$ as
\begin{equation}
  \left(
    \mathbf{B}_k(\boldsymbol{z};\tau)
  \right)_{ab}
  =
  B_k^{(b)}(z_a;\tau) ,
\end{equation}
for $1 \leq a,b \leq k$,
and $z_a\in \mathbb{C}$.
We introduce ~\cite{EguchiHikami08a}
\begin{equation}
  \label{define_H_k_z}
  H_k^{(a)}(z_1,\dots,z_k;\tau)
  =
  \sum_{b=1}^{k}
  \left(
    \mathbf{B}_k(\boldsymbol{z};\tau)^{-1}
  \right)_{ab} \,
  C_k(z_b;\tau) ,
\end{equation}
whose
completion is
\begin{equation}
  \label{define_H_hat}
  \widehat{H}_k^{(a)}(z_1,\dots,z_{k};\tau)
  =
  \sum_{b=1}^{k}
  \left(
    \mathbf{B}_k(\boldsymbol{z};\tau)^{-1}
  \right)_{ab} \,
  \widehat{C}_k(z_b;\tau) .
\end{equation}
We then have the modular transformation laws,
\begin{equation}
  \label{modular_H_hat}
  \begin{gathered}
    \widehat{H}_k^{(a)}(z_1,\dots,z_{k};\tau)
    =
    -\sqrt{\frac{\I}{\tau}}
    \sum_{b=1}^{k}
% \mathbf{S}(P)_{ab} \,
    \sqrt{\frac{2}{k+1}} \, \sin \left( \frac{a \, b}{k+1} \, \pi \right)
    \,
    \widehat{H}_k^{(b)}
    \left(
      \frac{z_1}{\tau}, \dots,
      \frac{z_{k}}{\tau} ;
      -\frac{1}{\tau}
    \right) ,
    \\[2mm]
    \widehat{H}_k^{(a)}(z_1,\dots,z_{k}; \tau+1)
    =
    \E^{- \frac{a^2}{2(k+1)} \, \pi \, \I} \,
    \widehat{H}_k^{(a)}(z_1,\dots,z_{k}; \tau) ,
    \\[2mm]
    \begin{aligned}
      \widehat{H}_k^{(a)}(z_1, \dots, z_b+1 , \dots, z_{k};\tau)
      & =
      \widehat{H}_k^{(a)}(z_1, \dots, z_b+\tau , \dots, z_{k};\tau)
      \\
      & =
      \widehat{H}_k^{(a)}(z_1, \dots,  z_{k};\tau) .
    \end{aligned}
  \end{gathered}
\end{equation}
We note that $\widehat{H}^{(a)}_k(z_i;\tau)$ and $H^{(a)}_k(z_i;\tau)$
are related as
\begin{equation}
  \widehat{H}^{(a)}_k(z_1,\cdots,z_k;\tau)
  =
  H^{(a)}_k(z_1,\cdots,z_k;\tau)-{R^{(a)}_k(\tau)\over \I\sqrt{2(k+1)}}.
  \label{H_hat_and_H}
\end{equation}

From~\eqref{define_R_k_a} it follows that
\begin{equation}
  \label{differential_R_k_a}
  \frac{\partial}{\partial \overline{\tau}} \,
  R_k^{(a)}(\tau)
  =
  \frac{
    \Psi_k^{(a)}(- \overline{\tau})
  }{
    \sqrt{2 \, \Im \tau}
  }  ,
\end{equation}
and we obtain
\begin{equation}
  \label{differential_H_Psi}
  \frac{\partial}{\partial \overline{\tau}}
  \widehat{H}_k^{(a)}(z_1,\dots,z_{k};\tau)
  =
  \frac{\I}{\sqrt{2 \, (k+1)}} \,
  \frac{1}{\sqrt{2 \, \Im \tau}} \,
  \Psi_k^{(a)} \left( - \overline{\tau} \right) .
\end{equation}
As a result, the completion
$\widehat{H}_k^{(a)}(z_1,\dots,z_{k};\tau)$ is a harmonic Maass
form, and 
is an eigenfunction of the differential operator
\begin{equation}
  \label{H_and_differential}
  \Delta_{\frac{1}{2}} \widehat{H}_k^{(a)}(z_1,\dots,z_{k};\tau)
  = 0 .
\end{equation}
Here $\Delta_{\ell}$ denotes the hyperbolic Laplacian ($\tau=u+\I \, v$) 
\begin{align}
  \Delta_{\ell}
  & =
  -v^2 \, \left( \frac{\partial^2}{\partial u^2}
    +
    \frac{\partial^2}{\partial v^2}
  \right)
  + \I \, {\ell}\, v \,
  \left(
    \frac{\partial}{\partial u}
    +
    \I \,    \frac{\partial}{\partial v}
  \right) 
  \nonumber \\
  & =
  -4 \,
  \left( \Im \tau \right)^{2-\ell}
  \frac{\partial}{\partial \tau}  
  \left( \Im \tau \right)^{\ell}
  \frac{\partial}{\partial \overline{\tau}} .
  \label{Laplacian}
\end{align}
Correspondingly, the function $H_k^{(a)}(z_1,\dots,z_{k};\tau)$
defined in~\eqref{define_H_k_z} is regarded as
a holomorphic part of the harmonic
Maass form.

\subsection{Jacobi Form}

The Jacobi form $f(z;\tau)$ with weight-$k$ and index-$m$ obeys the
following transformation laws~\cite{EichZagi85};
\begin{equation}
  \label{define_transform_Jacobi}
  \begin{gathered}
    f\left( \frac{z}{\tau} ; - \frac{1}{\tau} \right)
    =
    \tau^k \, \E^{2  \pi  \I  m  \frac{z^2}{\tau}} \,
    f(z ; \tau),
    \\[2mm]
    f(z;\tau+1) = f(z+1; \tau) = f(z; \tau),
    \\[2mm]
    f(z+\tau ; \tau)
    = q^{-m} \, \E^{- 4  \pi  \I  m  z} \, f(z;\tau) .
  \end{gathered}
\end{equation}
It is known~\cite{EichZagi85}
that the space of the Jacobi form with even weight
is spanned by
\begin{equation*}
  \left\{
    E_4(\tau),
    E_6(\tau),
    \phi_{-2,1}(z;\tau),
    \phi_{0,1}(z;\tau)
  \right\} ,
\end{equation*}
and 
a  basis of Jacobi forms with weight-$k$ and
index-$m$ is given by 
\begin{equation}
  \label{Eichler_Zagier_base}
  \left[ E_4(\tau) \right]^a \,
  \left[ E_6(\tau) \right]^b \,
  \left[ \phi_{-2,1}(z;\tau) \right]^c \,
  \left[ \phi_{0,1}(z;\tau) \right]^d,
\end{equation}
with non-negative integers
$a$, $b$, $c$, $d$ satisfying
\begin{align*}
  & 4 \, a + 6 \, b - 2 \, c = k, \hskip3mm 
   c+d=m .
\end{align*}
Here $E_4(\tau)$ and $E_6(\tau)$ are the Eisenstein series,
\begin{align*}
  E_4(\tau)
  & =
  1+ 240 \sum_{n=1}^\infty \sigma_3(n) \, q^n 
  \\
  & =
  1+ 240 \, q + 2160 \, q^2 +6720 \, q^3 +
  17520 \, q^4 + 30240 \, q^5 +
  \cdots,
  \\[2mm]
  E_6(\tau)
  & = 
  1 - 504 \sum_{n=1}^\infty \sigma_5(n) \, q^n
  \\
  & = 
  1 - 504 \, q - 16632 \, q^2 - 122976 \, q^3
  - 532728 \, q^4 - \cdots ,
\end{align*}
where
\begin{equation*}
  \sigma_k(n) = \sum_{r | n} r^k .
\end{equation*}
The remaining two functions with index-$1$ are defined by
\begin{align}
  \label{define_phi-21}
  {\phi}_{-2,1}(z;\tau)
  % & =
  % \frac{1}{
  %   \left[
  %     \eta(\tau)
  %   \right]^{24}
  % } \cdot
  % \frac{
  %   E_6 \, E_{4,1} - E_4 \, E_{6,1}
  % }{144}(z;\tau)
  % \\
  & =
  -
  \frac{
    \left[ \theta_{11}(z;\tau) \right]^2
  }{
    \left[ \eta(\tau) \right]^6
  } 
  \\
  & =
  \left(
    \zeta - 2 +\zeta^{-1}
  \right)
  +
  \left(
    -2  \, \zeta^2 +8 \, \zeta -12 +8 \, \zeta^{-1} -
    2 \, \zeta^{-2}
  \right) \, q
  \nonumber \\
  &  \quad \qquad
  +
  \left(
    \zeta^3 - 12 \, \zeta^2  +39 \, \zeta
    -56
    +39 \, \zeta^{-1} -12 \, \zeta^{-2} + \zeta^{-3}
  \right) \, q^2
  + \cdots   ,
  \nonumber
  \\[2mm]
  \label{define_phi01}
  {\phi}_{0,1}(z;\tau)
  % & =
  % \frac{1}{
  %   \left[
  %     \eta(\tau)
  %   \right]^{24}
  % } \cdot
  % \frac{
  %   E_4^{~2} \, E_{4,1} - E_6 \, E_{6,1}
  % }{144}(z;\tau)
  % \\
  & =
  4 \,
  \left[
    \left(
      \frac{\theta_{10}(z;\tau)}{\theta_{10}(0;\tau)}
    \right)^2
    +
    \left(
      \frac{\theta_{00}(z;\tau)}{\theta_{00}(0;\tau)}
    \right)^2
    +
    \left(
      \frac{\theta_{01}(z;\tau)}{\theta_{01}(0;\tau)}
    \right)^2
  \right] 
  \\
  & =
  \left(
    \zeta + 10 + \zeta^{-1}
  \right)
  +
  \left(
    10 \, \zeta^2 - 64 \, \zeta + 108 - 64 \, \zeta^{-1} +
    10 \, \zeta^{-2}
  \right) \, q
  \nonumber
  \\
  & \qquad \quad
  +
  \left(
    \zeta^3 + 108 \, \zeta^2 - 513 \, \zeta
    +
    808
    -513 \, \zeta^{-1} + 108 \, \zeta^{-2}
    + \zeta^{-3}
  \right) \, q^2
  + \cdots   .
  \nonumber
\end{align}
% where
% the
% Jacobi--Eisenstein series $E_{k,m}(z;\tau)$ is
% \begin{equation}
%   E_{k,m}(z;\tau)
%   =
%   \frac{1}{2}
%   \sum_{\substack{
%       c,d\in \mathbb{Z}
%       \\
%       (c,d)=1
%     }}
%   \sum_{\lambda \in \mathbb{Z}}
%   \left( c \, \tau + d \right)^{-k} \,
%   \E^{2 \pi \I m
%     \left(
%       \lambda^2 
%       \frac{a \tau+b}{c \tau+d}
%       + 2 \lambda \frac{z}{c \tau+d}
%       - \frac{c z^2}{c \tau+d}
%     \right)
%   } .
% \end{equation}
It is noted that
\begin{equation}
  \begin{aligned}
    \phi_{-2,1}(0;\tau) & = 0 ,
    \\[2mm]
    \phi_{0,1}(0;\tau) & = 12  ,
  \end{aligned}
\end{equation}
and that
$\phi_{0,1}(z;\tau)$ is just one-half of
the elliptic genus of the $K3$
surface~\cite{EguOogTaoYan89a,KawaYamaYang94a}.

%%%
\subsection{Character Decomposition of Elliptic Genera}
In terms of the completion $\widehat{C}_k(z;\tau)$ of the massless
character and the harmonic Maass form
$\widehat{H}^{(a)}_k(z_1, \dots, z_k;\tau)$,
we introduce
the function $J_k( z; w_1, \dots, w_{k};\tau )$
as ~\cite{EguchiHikami08a}
\begin{align}
  J_k \left( z; w_1,\dots,w_{k};\tau \right)
  & =
  \widehat{C}_k(z;\tau)
  -
  \sum_{a=1}^{k}
  \widehat{H}_k^{(a)} \left( w_1,\dots,w_{k};\tau \right) \,
  B_k^{(a)}(z;\tau)
  \label{J_decompose_hat_C_H}
  \\
  & =
  {C}_k(z;\tau)
  -
  \sum_{a=1}^{k}
  {H}_k^{(a)} \left( w_1,\dots,w_{k};\tau \right) \,
  B_k^{(a)}(z;\tau) .
  \label{J_decompose_C_H}
\end{align}
Non-holomorphic dependence in~\eqref{J_decompose_hat_C_H} cancels each
other,
and
the function $J_k(z;w_1,\dots,w_{k};\tau)$
is holomorphic as is seen in~\eqref{J_decompose_C_H}.
$J_k(z;w_1,\dots,w_{k};\tau)$ transforms like a 
Jacobi form with weight-$0$ and index-$k$~\cite{EichZagi85};
\begin{equation}
  \label{transform_J_k}
  \begin{gathered}
    J_k(z;w_1,\dots,w_{k};\tau)
    =
    \E^{-2 \pi \I k\frac{z^2}{\tau}} \,
    J_k\left(
      \frac{z}{\tau}; \frac{w_1}{\tau}, \dots, \frac{w_{k}}{\tau};
      -\frac{1}{\tau}
    \right) ,
    \\[2mm]
    \begin{aligned}
      J_k(z+1;w_1,\dots,w_{k};\tau)
      & =
      J_k(z;w_1,\dots,w_a+1,\dots,w_{k};\tau)
      \\
      & =
      J_k(z;w_1,\dots,w_a+\tau,\dots,w_{k};\tau)
      \\
      & =
      J_k(z;w_1,\dots,w_{k};\tau+1)
      \\
      &    =
      J_k(z;w_1,\dots,w_{k};\tau) ,
    \end{aligned} 
    \\[2mm]
    J_k(z+ \tau;w_1,\dots,w_{k};\tau)
    =
    q^{-k} \, \E^{-4 \pi \I k z} \,
    J_k(z;w_1,\dots,w_{k};\tau) .
  \end{gathered}
\end{equation}
By construction, the function $J_k(z; w_1,\dots,w_{k};\tau)$
vanishes at $z=w_a$ for $a=1,\dots,k$,
\begin{equation}
  J_k(w_a; w_1,\dots, w_{k};\tau) = 0 ,
\end{equation}
and we also have
\begin{equation}
  \label{J_at_0}
  J_k(0; w_1,\dots, w_{k};\tau) = 1 ,
\end{equation}
because of~\eqref{Witten_index}.
In the following we  choose $w_a$ to be half-periods,
$w_a \in
\left\{
 \frac{1}{2}, \frac{1+\tau}{2} , \frac{\tau}{2}
\right\}$,
and use the notation 
$\boldsymbol{w}_{(k_2,k_3,k_4)}$
\begin{multline*}
  \boldsymbol{w}_{(k_2,k_3,k_4)}
  =
  \\
  \left\{
    w_1,\dots,w_{k}
    ~ \Big| ~
    k_2 =
    \#  \left(
      w_a=\frac{1}{2}
    \right),
    k_3
    =
    \#
    \left(
      w_a=\frac{1+\tau}{2}
    \right),
    k_4
    =
    \#
    \left(
      w_a=\frac{\tau}{2}
    \right) 
  \right\} .
\end{multline*}
Then it is possible to show that 
\begin{equation}
  J_k(z; \boldsymbol{w}_{(k_2,k_3,k_4)} ; \tau)
  =
  \left(
    \frac{\theta_{10}(z;\tau)}{\theta_{10}(0;\tau)}
  \right)^{2 \, k_2}
  \,
  \left(
    \frac{\theta_{00}(z;\tau)}{\theta_{00}(0;\tau)}
  \right)^{2 \, k_3}
  \,
  \left(
    \frac{\theta_{01}(z;\tau)}{\theta_{01}(0;\tau)}
  \right)^{2 \, k_4} ,\hskip2mm k=k_2+k_3+k_4
\end{equation}
This is a vector-valued Jacobi form, and is a building block of the elliptic
genera for hyperK{\"a}hler
manifold with complex dimensions
$2k$~\cite{EgucSugaTaor07a,EgucSugaTaor08a}.
Symmetrization of $J_k(z;\boldsymbol{w}_{(k_2,k_3,k_4)};\tau)$ in
$k_2$,
$k_3$,
$k_4$ 
gives 
a Jacobi form with weight-$0$ and index-$k$.
%  and 
% it is a piece of the elliptic genus of hyperK{\"a}hler manifold
% $X_{P-1}$ with complex dimension $P-1$.

For our convenience we introduce the following notation, 
\begin{equation}
  \label{define_Sigma_k}
  \Sigma_{(k_2,k_3,k_4)}^{(a)}(\tau)
  =
  \sum_{%\mathclap{
      \text{symmetrization of $(k_2,k_3,k_4)$}}
  %}
  H_k^{(a)}
  \left(
    \boldsymbol{w}_{(k_2,k_3,k_4)} ; \tau
  \right) .
\end{equation}
Here $k_2+k_3+k_4=k$ and
without loss of generality we set
$k_2\geq k_3 \geq k_4$. 
Completion
$\widehat{\Sigma}_{(k_2,k_3,k_4)}^{(a)}(\tau)$
is defined as
\begin{equation}
  \widehat{\Sigma}_{(k_2,k_3,k_4)}^{(a)}(\tau)
  =
  \sum_{\text{symmetrization of $(k_2,k_3,k_4)$}}
  \widehat{H}_k^{(a)}
  \left(
    \boldsymbol{w}_{(k_2,k_3,k_4)} ; \tau
  \right) .
\end{equation}

As $C_k(z;\tau)$ and $B_k^{(a)}(z;\tau)$ are the massless and
massive characters~\eqref{define_C_k} and~\eqref{basis-function}
respectively,
the formula ~\eqref{J_decompose_C_H} is used to give the decomposition of
elliptic genera in terms of $\mathcal{N}=4$ irreducible representations.
In particular, the Fourier coefficients of 
${H}_k^{(a)}(\boldsymbol{w}_{(k_2,k_3,k_4)};\tau)$ counts 
the
number of massive representations in elliptic genera.
Since in
elliptic genera the right-moving sectors are  always fixed to the
ground state, massive representations in the left-moving sector
correspond to the overall half-BPS. 
Then the asymptotic behavior of the
growth of the multiplicity of
non-BPS states in elliptic genera is related to the black
hole entropy in string compactification on hyperK\"ahler manifolds.
As
we shall see in the standard case of D5-D1 black hole in string
compactification on $K3$ surface, we  
will reproduce the black hole entropy from the growth of
massive representations.

%%%%%%%%%%%%%%%5
\section{Harmonic Maass Form and Poincar{\'e}--Maass Series}
\label{sec:Poincare}

\subsection{Jacobi Form and Theta Series}
\label{sec:integrality}
In the formula ~\eqref{J_decompose_C_H},
the Fourier coefficients of $H_k^{(a)}(w_1,\dots,w_{k};\tau)$ count 
the multiplicity of non-BPS representations. 
Our purpose is to compute these Fourier coefficients.
As the parameters $w_a$ are specialized to half-period,
our problem is to construct
a vector-valued harmonic Maass form ($k \geq 1$ and
$1 \leq a \leq k$),
\begin{equation}
  \label{Laplacian_Poincare}
  \Delta_{\frac{1}{2}} \widehat{\Sigma}_k^{(a)}(\tau) = 0,
\end{equation}
which 
transforms as ~\eqref{modular_H_hat};
\begin{equation}
  \label{ST_transform_Poincare}
  \begin{gathered}
    \widehat{\Sigma}_k^{(a)}(\tau)
    = 
    - \sqrt{\frac{\I}{\tau}} \,
    \sum_{b=1}^{k}
    \sqrt{\frac{2}{k+1}} \,
    \sin \left(
      \frac{a \, b}{k+1} \, \pi
    \right) \,
    \widehat{\Sigma}_k^{(b)} \left( - \frac{1}{\tau} \right) ,
    \\[2mm]
    \widehat{\Sigma}_k^{(a)}(\tau+1)
    =
    \E^{ - \frac{a^2}{2 \, (k+1)} \,  \pi \, \I} \,
    \widehat{\Sigma}_k^{(a)}(\tau) .
  \end{gathered}
\end{equation}

Once we are given such a modular form, we can construct a real
analytic Jacobi form 
$\widehat{\mathcal{J}}_k(z;\tau)$ of weight-$0$ 
and index-$k$ by 
\begin{equation}
  \label{Jacobi_decompose_B}
  \widehat{\mathcal{J}}_k(z;\tau)
  =
  \sum_{a=1}^{k} \widehat{\Sigma}_k^{(a)}(\tau) \, B_k^{(a)}(z;\tau).
\end{equation}
Note that,
when $\widehat{\Sigma}_k^{(a)}(\tau)$ is holomorphic
and trivially  satisfies~\eqref{Laplacian_Poincare},
the function
$\widehat{\mathcal{J}}_k(z;\tau)$ becomes a holomorphic Jacobi form.

On the contrary, we can invert the above relation and determine the
function $\widehat{\Sigma}_k^{(a)}(\tau)$ 
in terms of  a real analytic Jacobi form
$\widehat{\mathcal{J}}_k(z;\tau)$. 
If we introduce a function 
\begin{equation*}
  \widetilde{\mathcal{J}}_k(z;\tau)
  =
  -\I \,
  \left[ \eta(\tau) \right]^3
  \,
  \frac{
    \theta_{11}(2 \, z ; \tau)
  }{
    \left[ \theta_{11}(z;\tau) \right]^2
  } \,
  \widehat{\mathcal{J}}_k(z;\tau)
\end{equation*}
for convenience,
which is a real analytic
Jacobi form with weight-$1$ and index-$(k+1)$,
we can in fact express  
the function $\widehat{\Sigma}_k^{(a)}(\tau)$ as a Fourier integral
\begin{equation}
  \label{h_Fourier_J}
  \widehat{\Sigma}_k^{(a)}(\tau)
  =
  q^{- \frac{a^2}{4(k+1)}} \,
  \int_{z_0}^{z_0+1}
   \,
  \widetilde{\mathcal{J}}_k(z;\tau) \, \E^{-2  \pi  \I   a  z} \,
  \mathrm{d}z ,
\end{equation}
where $z_0\in \mathbb{C}$ is arbitrary.
Proof of~\eqref{h_Fourier_J} is rather standard~\cite{EichZagi85};
due to the periodicity of $\widetilde{\mathcal{J}}_k(z;\tau)$ in
$z\to z +1$,
% in $\mathbb{Z}$,
we can expand
\begin{equation*}
  \widetilde{\mathcal{J}}_k(z;\tau)
  =
  \sum_{n \in \mathbb{Z}}
  q^{\frac{n^2}{4(k+1)}} \,
 \widehat{\Sigma}_{k}^n(\tau) \, \E^{2  \pi  \I  n  z} ,
\end{equation*}
where $q^{\frac{n^2}{4(k+1)}}$ is inserted for  convenience.
Coefficients $\widehat{\Sigma}_{k}^{n}(\tau)$ are given by
\begin{equation*}
  \widehat{\Sigma}_{k}^{n}(\tau)
  =
  q^{-\frac{n^2}{4(k+1)}} \,
  \int_{z_0}^{z_0+1} \widetilde{\mathcal{J}}_k(z;\tau) \,
  \E^{-2  \pi  \I  n  z} \,
  \mathrm{d} z .
\end{equation*}
Quasi-periodicity of 
$\widetilde{\mathcal{J}}_k(z;\tau)$ in $z\to z +\tau$
% in $\mathbb{Z} \, \tau$
implies
$\widehat{\Sigma}_{k}^{n}(\tau) =\widehat{\Sigma}_{k}^{ n+2(k+1)}(\tau)$.
We thus obtain
\begin{align*}
  \widetilde{\mathcal{J}}_k(z;\tau)
  & =
  \sum_{m \in \mathbb{Z}}\hskip1mm  \sum_{a \hskip-2mm \mod 2(k+1)}
  \widehat{\Sigma}_{k}^{a}(\tau) \, \E^{2  \pi  \I  (2(k+1)m + a)  z} \,
  q^{\frac{
      (2(k+1)m + a)^2}{4(k+1)}
  }
  \\
  & =
  \sum_{a\hskip-2mm \mod 2(k+1)}
  \widehat{\Sigma}_{k}^{a}(\tau) \, \vartheta_{k+1,a}(z;\tau).
\end{align*}
Since $\widetilde{\mathcal{J}}_k(z;\tau)$ is odd  with respect to $z$
and $\vartheta_{k+1,a}(-z;\tau)=\vartheta_{k+1,-a}(z;\tau)$,
we recover~\eqref{Jacobi_decompose_B}.

% In case of~\eqref{J_decompose_C_H}
% $J_P(z;\w_1,\dots,w_{P-1};\tau) - \widehat{C}_P(z;\tau)$
% plays a role of $\widehat{J}_P(z;\tau)$.
In the case when 
$\widehat{\mathcal{J}}_k(z;\tau)$ is real analytic, for example
$J_k(z;w_1,\dots,w_{k};\tau)-\widehat{C}_k(z;\tau)$
as in~\eqref{J_decompose_hat_C_H},
formula ~\eqref{h_Fourier_J} is valid 
when we  replace $\widehat{\mathcal{J}}_k(z;\tau)$
with
$J_k(z;w_1,\dots,w_{k};\tau)-\widehat{C}_k(z;\tau)$.
It is possible to see that also in the holomorphic case the
relation~\eqref{h_Fourier_J} is valid 
when we replace $\widehat{\Sigma}^{(a)}_k(\tau)$ by
${\Sigma}^{(a)}_k(\tau)$
and $\widehat{\mathcal{J}}_k(z;\tau)$ by 
$J_k(z;w_1,\dots,w_{k};\tau)-{C}_k(z;\tau)$.
This is due to the
relationship~\eqref{general_Maass_pre} and (\ref{H_hat_and_H}).
%and~\eqref{define_H_k_z}--\eqref{define_H_hat}, which give
%   \widehat{C}_k(z;\tau)
%   =
%   C_k(z;\tau)
%   -
%   {1\over \I \sqrt{2 \, (k+1)}}
%   \sum_{a=1}^k R_k^{(a)}(\tau) \, B_k^{(a)}(z;\tau),
%   \\

% Furthermore  the holomorphic part of $\widehat{C}_P(z;\tau)$ gives
% \begin{equation*}
%   -\I \,
%   \left[ \eta(\tau) \right]^3
%   \,
%   \frac{
%     \theta_{11}(2 \, z ; \tau)
%   }{
%     \left[ \theta_{11}(z;\tau) \right]^2
%   } \,
%   C_P(z;\tau)
%   =
%   \sum_{n \in \mathbb{Z}}
%   q^{P n^2} \, \E^{4 \pi \I P n z} \,
%   \frac{
%     1+ q^n \, \E^{2 \pi \I z}
%   }{
%     1- q^n \, \E^{2 \pi \I z}
%   } ,
% \end{equation*}
% which contains
% $\E^{2 \, \pi \, \I \, m \, z}$ with $|m|\geq 2 \, P$ for
% $q^{\mathbb{Z}_{>0}}$,
% the holomorphic function $h_P^{(a)}(\tau)$ with $1\leq a \leq P-1$ is given
% up to a $q^0$ term
% by~\eqref{h_Fourier_J} replacing
% $\widehat{J}_P(z;\tau)$ with
% $J_P(z;w_1, \dots, w_{P-1};\tau)$.

Using the fact that
\begin{multline*}
  - \I \, \theta_{11}(2 \, z; \tau) \,
  \frac{
    \left[ \eta(\tau) \right]^3}{
    \left[
      \theta_{11}(z;\tau)
    \right]^2
  }
  \\
  =
  \frac{1+\zeta}{1-\zeta}
  +
  \left( \zeta^2 - \zeta^{-2} \right) \, q
  +
  2 \,
  \left( \zeta^3 - \zeta^{-3} \right) \, q^2
  +
  2 \,
  \left( \zeta^4 - \zeta^{-4} \right) \, q^3
  \\
  +
  \left(
    2 \, \zeta^5 + \zeta^4-\zeta^{-4}-2 \, \zeta^{-5}
  \right) \,  q^4
  +
  2 \,
  \left(
    \zeta^6 - \zeta^{-6}
  \right) \, q^5
  + \cdots ,
\end{multline*}
integrality of the Fourier coefficients of
$\widehat{\Sigma}_k^{(a)}(\tau)$ in~\eqref{h_Fourier_J}
follows straightforwardly
once one has  integrality of the Fourier coefficients of
the Jacobi form $\widehat{\mathcal{J}}_k(z;\tau)$.

In the case of $k=1$ we take the 
Jacobi form to be the elliptic genus of the $K3$ surface
$2 \, \phi_{0,1}(z;\tau)$.
Then we find 
\begin{multline*}
  -\I \,
  \left[ \eta(\tau) \right]^3
  \,
  \frac{
    \theta_{11}(2 \, z ; \tau)
  }{
    \left[ \theta_{11}(z;\tau) \right]^2
  } \,
  \left[
    2 \, \phi_{0,1}(z;\tau)
    -
    24 \, C_1(z;\tau)
  \right] 
  \\
  =
  \left( -2 \, \zeta  + 2 \, \zeta^{-1} \right)
  +
  \left( 2 \, \zeta^3 + 90 \, \zeta
    - 90 \, \zeta^{-1} -2 \, \zeta^{-3} \right) \, q
  \\
  +
  \left(
    -90 \, \zeta^3 + 462 \, \zeta
    - 462 \, \zeta^{-1} + 90 \, \zeta^{-3}
  \right) \, q^2
  \\
  +
  \left(
    -2 \, \zeta^5 - 462 \, \zeta^{3} + 1540 \, \zeta
    - 1540 \, \zeta^{-1}  + 462 \, \zeta^{-3} + 2 \, \zeta^{-5}
  \right) \, q^3
  \\
  +
  \left( 
    90 \, \zeta^5 - 1540 \, \zeta^3  + 4554 \, \zeta
    - 4554 \, \zeta^{-1} + 1540 \, \zeta^{-3} - 90 \, \zeta^{-5}
  \right) \, q^4 + \cdots .
%   \frac{2 \, (1 + \zeta) \,(10+\zeta+\zeta^{-1})}{
%     1 - \zeta}
%   + \left(
%     2 \, \zeta^3 + 90 \, \zeta - 90 \, \zeta^{-1}-2 \, \zeta^{-3}
%   \right)
%   \, q
%   \\
%   +
%   \left(
%     24 \, \zeta^4 - 90 \, \zeta^3+462 \, \zeta - 462 \, \zeta^{-1}
%     + 90 \, \zeta^{-3} - 24 \, \zeta^{-4}
%   \right) \, q^2
%   \\
%   +
%   \left(
%     46 \, \zeta^5-462 \, \zeta^3 + 1540 \, \zeta - 1540 \, \zeta^{-1}
%     + 462 \, \zeta^{-3} - 46 \, \zeta^{-5}
%   \right) \, q^3
%   \\
%   +
%   \left(
%     48 \, \zeta^6+90 \, \zeta^5
%     - 1540 \, \zeta^3 + 4554 \, \zeta
%     - 4554 \, \zeta^{-1} + 1540 \, \zeta^{-3}
%     -90 \, \zeta^{-5} - 48 \, \zeta^{-6}
%   \right) \, q^4
%   + \cdots .
\end{multline*}
One finds that the Fourier coefficients of $\zeta$ or $\zeta^{-1}$ 
are
nothing but the multiplicity of massive representations in the $K3$
surface discussed in~\cite{EguchiHikami09a}.

\subsection{Multiplier System}
We shall  construct a solution of~\eqref{Laplacian_Poincare}
and~\eqref{ST_transform_Poincare} in the form of the
Poincar{\'e}--Maass series.
It is a generalization of the discussion in 
our previous
paper~\cite{EguchiHikami09a} where
a case of $k=1$ was studied as an application of the Rademacher
expansion for the mock theta
function. 
See Refs.~\citenum{BrinKOno06a,BrinKOno08a}
for recent studies on the Poincar{\'e}--Maass series.

We utilize   the following multiplier system for the SU(2) affine
character~\eqref{define_affine_SU2}.
For
$\gamma=
\begin{pmatrix}
  a & b \\
  c & d
\end{pmatrix}
\in \Gamma(1)=SL(2;\mathbb{Z})$,
we set
\begin{equation}
  % \frac{
  %   \Psi_P^{(\ell_1)}}{
  %   \eta^3}
  % \left( \gamma(\tau) \right)
  \chi_{k-1,\frac{a_1 -1}{2}}\left( 0;\gamma(\tau) \right)
  =
  \sum_{a_2= 1}^{k}
  \left[
    \rho(\gamma)
  \right]_{a_1,a_2} \,
  \chi_{k-1,\frac{a_2 -1}{2}}(0;\tau) .
  % \frac{
  %   \Psi_P^{(\ell_2)}}{
  %   \eta^3}
  % (\tau)
\end{equation}
Here we have
\begin{equation}
  \begin{aligned}
    \left[
      \rho
      \left(
        \begin{pmatrix}
          0 & -1 \\
          1 & 0
        \end{pmatrix}
      \right)
    \right]_{a_1,a_2}
    & =
    \sqrt{\frac{2}{k+1}} \,
    \sin
    \left(
      \frac{
        a_1  \,  a_2 }{k+1} \,
      \pi
    \right),
    \\[2mm]
    \left[
      \rho
      \left(
        \begin{pmatrix}
          1 & 1 \\
          0 & 1
        \end{pmatrix}
      \right)
    \right]_{a_1, a_2}
    & =
    \E^{
      \left(
        \frac{a_1^{~2}}{2 (k+1)} - \frac{1}{4}
      \right) \,
      \pi \, \I
    } \,
    \delta_{a_1, a_2} ,
  \end{aligned}
  \label{S_T_chi}
\end{equation}
and, 
in general
(see, \emph{e.g.}, Refs.~\citenum{LCJeff92a,SkorZagi89a})
\begin{multline}
  \left[
    \rho
    (\gamma)
  \right]_{a_1, a_2}
  =
  -\I \frac{\sign(c)}{\sqrt{2 (k+1)
      \left| c \right|
    }}
  \, \E^{-\frac{a+d}{4 \,c} \, \pi \, \I
    +
    3 \, s(d,c) \, \pi \, \I
  }
  \,
  \E^{\frac{d \, a_2^{~2}}{2 \, (k+1) \, c} \, \pi \, \I}
  \\
  \times
  \sum_{\substack{
      j=0 \\
      j \equiv a_1 \mod 2(k+1)
    }}^{2(k+1)c -1}
  \E^{\frac{ a \, j^2}{2(k+1) c} \, \pi \, \I} \,
  \left(
    \E^{\frac{a_2 \, j}{(k+1) \, c} \, \pi \, \I}
    -
    \E^{-\frac{a_2 \, j}{(k+1) \, c} \, \pi \, \I}
  \right) .
\end{multline}
Here $s(d,c)$ is the Dedekind sum defined by
\begin{equation*}
  s(d,c)
  =
  \sum_{k \mod c}
  \Biggl( \!\!  \Biggl( \frac{k}{c} \Biggr) \!\! \Biggr) \,
  \Biggl( \!\!  \Biggl( \frac{k\, d}{c} \Biggr) \!\! \Biggr) ,
\end{equation*}
where
\begin{equation*}
  ( \! ( x ) \! ) =
  \begin{cases}
    \displaystyle
    x - \lfloor x \rfloor - \frac{1}{2}  ,
    &
    \text{for $ x \in \mathbb{R} \setminus \mathbb{Z}$,}
    \\[2mm]
    0 , &
    \text{for $x \in \mathbb{Z}$.}
  \end{cases}
\end{equation*}
This  representation has been used~\cite{LCJeff92a}
to construct the SU(2)
Witten--Reshetikhin--Turaev invariant
of 3-manifold~\cite{EWitt89a,ResheTurae91a} from the colored Jones
polynomial for link to be  surgered.

Based on the similarity between
the modular transformations~\eqref{ST_transform_Poincare}
and~\eqref{S_T_chi},
the multiplier system for $\widehat{\Sigma}_k^{(a)}(\tau)$ can be given
explicitly.
Making use of the modular transformation  for the Dedekind
$\eta$-function
($c>0$; see, \emph{e.g.}, Ref.~\citenum{HRadema73}),
\begin{equation}
  \eta \left( \gamma(\tau) \right)
  =
  \I^{-\frac{1}{2}} \,
  \E^{\frac{a+d}{12 \, c} \, \pi \, \I
    -s(d,c) \, \pi \, \I} \,
  \sqrt{c \, \tau+d} \,
  \eta(\tau),
\end{equation}
we have the multiplier system for the vector-valued modular
form~\eqref{ST_transform_Poincare} as
\begin{equation}
  \label{general_transform_Poincare}
  \widehat{\Sigma}_{k}^{(a_1)}
  \left(
    \gamma(\tau)
  \right)
  =
  \sqrt{c \, \tau + d}
  \sum_{a_2=1}^{k}
  \left[
    \chi(\gamma)
  \right]_{a_1,  a_2} \,
  \widehat{\Sigma}_k^{(a_2)}(\tau) ,
\end{equation}
where
\begin{multline}
  \label{define_chi_a}
  \left[
    \chi(\gamma)
  \right]_{a_1, a_2}
  \\
  =
  \begin{cases}
    \displaystyle
    \sqrt{\I} \, \frac{
      \sign(c)}{\sqrt{
        2 (k+1) 
        \left| c \right|
      }} \,
    \E^{- \frac{a_2^{~2}}{2(k+1)} \, \frac{d}{c} \, \pi \, \I}
    \sum_{\substack{
        j=0\\
        j \equiv a_1 \mod 2(k+1)
      }}^{2(k+1)|c| - 1}
    \E^{- \frac{a \, j^2}{2(k+1)c} \, \pi \, \I} \,
    \left(
      \E^{\frac{a_2 \, j}{(k+1)c} \, \pi \, \I}
      -
      \E^{- \frac{a_2 \, j}{(k+1)c} \, \pi \, \I}
    \right) ,
    & \text{for $c \neq 0 $},
    \\[2mm]
    \displaystyle
    \delta_{a_1, a_2} \,
    \E^{- \frac{a_1^{~2} \, b}{2(k+1)} \, \pi \, \I
    } ,
    &    \text{for $c=0$.}
  \end{cases}
\end{multline}

%%%%%%
\subsection{Poincar{\'e}--Maass Series}
We shall construct  the harmonic Maass form in the form
of the Poincar{\'e}--Maass series $P_k^{(a)}(\tau)$.
We suppose that
the holomorphic  polar part of $P_k^{(a)}(\tau)$  has a form of
\begin{equation}
  \label{def_polar}
  \left.
    P_k^{(a)}(\tau)
  \right|_{\text{polar}}
  =
  \sum_{0 \leq n < \frac{a^2}{4(k+1)}} p_k^{(a)}(n) \, q^{n- \frac{a^2}{4(k+1)}} .
\end{equation}
Following Ref.~\citenum{Bruin02Book},
we set
for $h>0$
\begin{equation}
  \varphi_{-h,s}^{\ell}(\tau)
  =
  \mathcal{M}_s^{\ell}
  \left( -4 \, \pi \, h \, \Im (\tau) \right) \,
  \E^{- 2 \,  \pi \,  \I \,  h  \, \Re(\tau)} .
\end{equation}
Here the function $\mathcal{M}_s^{\ell}(v)$ is defined by
\begin{equation*}
  \mathcal{M}_s^{\ell}(v)
  =
  \left| v \right|^{- \frac{\ell}{2}} \,
  M_{\frac{\ell}{2} \sign(v), s - \frac{1}{2}}
  \left( \left| v \right| \right) ,
\end{equation*}
where
$M_{\alpha,\beta}(z)$ is the $M$-Whittaker function~\cite{WhittWatso27}.
We see that the $\varphi$-function is an eigenfunction of the
hyperbolic Laplacian~\eqref{Laplacian}
\begin{equation}
  \Delta_{\ell} \, \varphi_{-h,s}^{\ell}(\tau)
  =
  \left[
    s \, (1-s) + \frac{\ell}{2} \, \left( \frac{\ell}{2}-1 \right)
  \right] \,
  \varphi_{-h,s}^{\ell}(\tau) ,
\end{equation}
and that
at $\Im \tau\to +\infty$
\begin{equation*}
  \varphi^{\ell}_{-h,s}(\tau)
  \sim
  \frac{\Gamma (2 \, s)}{
    \Gamma
    \left(
      \frac{\ell}{2}+s
    \right)
  } \,
  q^{-h} .
\end{equation*}

By use of the Fourier coefficients of the polar 
part~\eqref{def_polar},
we  construct the  Poincar{\'e}--Maass series $P_k^{(a_1)}(\tau)$ for
$k \geq 1$ and $1 \leq a_1 \leq k$ by
\begin{multline}
  \label{define_Poincare-Maass}
  P_{k}^{(a_1)}(\tau)
  =
  \frac{1}{\sqrt{\pi}}
  \sum_{a_2=1}^{k}
  \sum_{0 \leq m < \frac{a_2^{~2}}{4(k+1)}}
  p_k^{(a_2)}(m)
  \\
  \times
  \sum_{
    \gamma=
    {\scriptsize
      \begin{pmatrix}
        a & b\\
        c & d
      \end{pmatrix}
    }
    \in \Gamma_\infty \backslash \Gamma(1)
  }
  \left[
    \chi \left(
      \gamma^{-1}
    \right)
  \right]_{a_1, a_2} \,
  \frac{1}{\sqrt{c \, \tau+d}} \,
%  \varphi_{-\Delta(\ell_2), \frac{3}{4}}^{\frac{1}{2}}
  \varphi_{m-\frac{a_2^{~2}}{4(k+1)}, \frac{3}{4}}^{\frac{1}{2}}
  \left(
    \gamma(\tau)
  \right) .
\end{multline}
Here
$\Gamma_\infty$ is the stabilizer of $\infty$,
\begin{equation*}
  \Gamma_\infty =
  \left\{
    \begin{pmatrix}
      1 & n \\
      0 & 1
    \end{pmatrix}
%    ~\middle|~
    ~\Big|~
    n \in \mathbb{Z}
  \right\} .
\end{equation*}
Commutativity of the Laplacian~\eqref{Laplacian}
and the $\gamma$-action proves
that
the Poincar{\'e}--Maass series satisfies~\eqref{Laplacian_Poincare},
and 
we can check that it fulfills
the modular transformation~\eqref{general_transform_Poincare}.

The Fourier coefficients of the Poincar{\'e}--Maass series
can be computed by the method developed in
Refs.~\citenum{BrinKOno06a,BrinKOno08a}
(see also Ref.~\citenum{EguchiHikami09a}).
We can rewrite $P_k^{(a_1)}(\tau)$ as
\begin{multline*}
  P_k^{(a_1)}(\tau)
  =
  \frac{2}{\sqrt{\pi}} 
  \sum_{0 \leq m < \frac{a_1^{~2}}{4(k+1)}}
  p_k^{(a_1)}(m)
  \, \varphi_{m-\frac{a_1^{~2}}{4(k+1)},
    \frac{3}{4}}^{\frac{1}{2}}(\tau)
  \\
  +
  \frac{1}{\sqrt{\pi}}
  \sum_{a_2}
  \sum_{0 \leq m < \frac{a_2^{~2}}{4(k+1)}}
  p_k^{(a_2)}(m)
  \sum_{\substack{
      c \neq 0
      \\
      \gamma \in \Gamma_\infty \backslash \Gamma(1) / \Gamma_\infty
    }}
  \sum_{n \in \mathbb{Z}}
  \left[
    \chi\left(
      \left( \gamma \,
        \begin{pmatrix}
          1 & n
          \\
          0 & 1
        \end{pmatrix}
      \right)^{-1}
    \right)
  \right]_{a_1, a_2}
  \\
  \times
  \frac{1}{
    \sqrt{ c \, \left( \tau+ n \right) + d}
  } \,
  \varphi_{m-\frac{a_2^{~2}}{4(k+1)},
    \frac{3}{4}}^{\frac{1}{2}}
  \left(
    \gamma \,
    \begin{pmatrix}
      1 & n \\
      0 & 1 
    \end{pmatrix}
    (\tau)
  \right) .
\end{multline*}
The second term reads up to a constant as
\begin{multline*}
  \sum_{a_2}
  \sum_{0 \leq m < \frac{a_2^{~2}}{4(k+1)}} p_k^{(a_2)}(m)
  \sum_{c >0}
  \frac{1}{\sqrt{c}}
  \sum_{\gamma \in \Gamma_\infty \backslash \Gamma(1) / \Gamma_\infty}
  \left[
    \chi \left( \gamma^{-1} \right) 
  \right]_{a_1, a_2} \,
  \E^{-2 \, \pi \, \I \,
    \left(
      \frac{a_2^{~2}}{4(k+1)}
      -m
    \right)
    \, \frac{a}{c}}
  \\
  \times
  \sum_{n \in \mathbb{Z}}
  \frac{1}{\sqrt{ \tau + n+ \frac{d}{c}}} \,
  \mathcal{M}_{\frac{3}{4}}^{\frac{1}{2}}
  \left(
    - 4 \, \pi \,
    \left(
      \frac{a_2^{~2}}{4(k+1)} -m
    \right) \,
    \frac{\Im \tau}{
      c^2 \,
      \left|
        \tau+ n  + \frac{d}{c}
       \right|^2
     }
   \right)
   \\
   \times
   \E^{2 \, \pi \, \I \, \frac{a_1^{~2}}{4(k+1) } \, n
     + 2 \, \pi \, \I \,
     \left(
       \frac{a_2^{~2}}{4(k+1) }
       -m
     \right) \, 
     \frac{1}{c^2} \,
     \Re
     \left(
       \frac{1}{\tau+n+\frac{d}{c}}
     \right)
   } .
\end{multline*}
We then apply the following Fourier transformation
formula~\cite{Bruin02Book,Hejhal83Book},
\begin{multline}
  \sum_{n \in \mathbb{Z}}
  \frac{1}{
    \sqrt{\tau +n}} \,
  \mathcal{M}^{\frac{1}{2}}_s \left(
    -4 \, \pi \, h \, 
    \frac{\Im (\tau) }{
      c^2 \, \left| \tau + n \right|^2}
  \right) \,
  \E^{2 \, \pi \,  \I \,  h^\prime \, n
    +2 \,  \pi \,  \I \,  \frac{h}{c^2}
    \Re \left( \frac{1}{\tau+n} \right)
  }
  \\
  =\sum_{n \in \mathbb{Z}} \,
  a_n\left(\Im (\tau) \right) \,
  \E^{2 \, \pi \,  \I \, 
    \left( n-h^\prime \right)
    \, \Re(\tau)
  } ,
\end{multline}
where the Fourier coefficients $a_n(v)$  are given as follows;
\begin{itemize}
\item for $n>h^\prime$,
  \begin{multline*}
    a_n(v)
    =
    \frac{1}{\sqrt{\I}} \,
    \left( \frac{h}{4 \, \pi \, c^2 \, v} \right)^{\frac{1}{4}}
    \,
    \frac{
      \Gamma(2 \, s)}{
      \Gamma \left( s+\frac{1}{4} \right)
    }
    \\
    \times
    \frac{2 \, \pi 
    }{
      \sqrt{n-h^\prime}} \,
    W_{\frac{1}{4}, s - \frac{1}{2}}
    \left(
      4 \, \pi \,
      \left( n - h^\prime \right) \, v 
    \right) \,
    I_{2s-1}
    \left(
      \frac{4 \, \pi}{
        \left| c \right|} \,
      \sqrt{\left( n-h^\prime \right) \, h}
    \right) ,
  \end{multline*}

\item for $n=h^\prime$,
  \begin{equation*}
    a_n(v)
    =
    \frac{1}{\sqrt{\I}} \,
    \frac{
      2^{\frac{3}{2}} \, \pi^{s+\frac{3}{4}} \,
      \Gamma(2 \, s)
    }{
      (2 \, s -1) \,
      \Gamma \left(s+\frac{1}{4} \right) \,
      \Gamma \left(s-\frac{1}{4} \right)
    } \,
    \frac{h^{s-\frac{1}{4}}}{
      \left| c \right|^{2s - \frac{1}{2}} \,
      v^{s-\frac{3}{4}}
    } ,
  \end{equation*}

\item for $n<h^\prime$,
  \begin{multline*}
    a_n(v)
    =
    \frac{1}{\sqrt{\I}} \,
    \left( \frac{h}{4 \, \pi \, c^2 \, v} \right)^{\frac{1}{4}}
    \,
    \frac{
      \Gamma(2 \, s)}{
      \Gamma \left( s - \frac{1}{4} \right)
    }
    \\
    \times
    \frac{2 \, \pi
    }{
      \sqrt{h^\prime - n}} \,
    W_{- \frac{1}{4}, s - \frac{1}{2}}
    \left(
      4 \, \pi \,
      \left(  h^\prime - n \right) \, v 
    \right) \,
    J_{2s-1}
    \left(
      \frac{4 \, \pi}{
        \left| c \right|} \,
      \sqrt{\left( h^\prime - n \right) \, h}
    \right) .
  \end{multline*}
\end{itemize}
Here 
%the $I$-Bessel
the (modified) Bessel function,  $I_\alpha(z)$ and $J_\alpha(z)$,
satisfy
\begin{gather*}
  \begin{aligned}
    I_\alpha(z)
    &\underset{z\to 0}{\sim} \frac{1}{
      \Gamma(\alpha+1)}
    \left( \frac{z}{2} \right)^{\alpha} \,
     ,
    &
    I_\alpha(z)
    & \underset{
      \left|z \right| \to \infty
    }{\sim}
    \frac{1}{
      \sqrt{2 \, \pi \, z}} \,
    \E^z ,
    \\
    J_\alpha(z)
    &\underset{z\to 0}{\sim}\frac{1}{
      \Gamma(\alpha+1)}
    \left( \frac{z}{2} \right)^{\alpha} \,
     ,
    & 
    J_\alpha(z)
    & \underset{
      \left|z \right| \to \infty
    }{\sim}
    \sqrt{\frac{2}{\pi \, z}} \,
    \cos
    \left(
      z- \frac{\alpha}{2} \, \pi - \frac{1}{4} \, \pi
    \right) .
  \end{aligned}
\end{gather*}
Substituting the above Fourier transformation formula,
we obtain  the expansion coefficients of the Poincar{\'e}--Maass
series as 
\begin{multline}
% \widehat{\Sigma}^{(a_1)}_k(\tau)=
  P_{k}^{(a_1)}(\tau)
  =
  \frac{2}{\sqrt{\pi}} \,
  \sum_{0 \leq m < \frac{a_1^{~2}}{4(k+1)}}
  p_k^{(a_1)}(m) \,
  \varphi_{m- \frac{a_1^{~2}}{4(k+1)}, \frac{3}{4}}^{\frac{1}{2}}(\tau)
  \\
  +
  \sum_{\substack{
      n \in \mathbb{Z}
      \\
      n
      \geq
      \frac{a_1^{~2}}{4 \, (k+1)}}}
  q^{n - \frac{a_1^{~2}}{4(k+1)}}
  \sum_{a_2 =1}^{k}
  \sum_{0 \leq m < \frac{a_2^{~2}}{4(k+1)}} p_k^{(a_2)}(m) 
  \sum_{\substack{c>0
      \\
      \gamma\in \Gamma_\infty \backslash \Gamma(1) / \Gamma_\infty
    }}
  \left[
    \chi \left(
      \gamma^{-1}
    \right)
  \right]_{a_1, a_2} \,
  \frac{2 \,  \pi}{\sqrt{\I}}
  \,
  \left[
    \frac{a_2^{~2}- 4(k+1)m}{4(k+1)n - a_1^{~2}}
  \right]^{\frac{1}{4}}
  % \left[
  %   \frac{\Delta(\ell_2)}{n - \frac{\ell_1^2}{4 \, P}}
  % \right]^{\frac{1}{4}}
  \\
  \times
  \frac{1}{c} \, 
  I_{\frac{1}{2}}
  \left(
    \frac{4 \, \pi}{c} \,
    \sqrt{
      \left(
        n - \frac{a_1^{~2}}{4(k+1)}
      \right) \,
%      \Delta(\ell_2)
      \left(
        \frac{a_2^{~2}}{4(k+1)} 
        -
        m
      \right)
    }
  \right) \, 
  \E^{-2 \,  \pi \,  \I \,
    \left(
      \frac{a_2^{~2}}{4(k+1)}
      - m
    \right) \,
%\Delta(\ell_2)
    \frac{a}{c} + 2 \, \pi \, \I \,
    \left(
      n- \frac{a_1^{~2}}{4(k+1)}
    \right) \, 
    \frac{d}{c}
  }
  \\
%   + \delta_{\frac{\ell_1^{~2}}{4 \, P} \in \mathbb{Z}}
%   \sum_{\ell_2=1}^{P-1} \sum_{0 \leq m < \frac{\ell_2^{~2}}{4 \, P}}
%   p_P^{(\ell_2)}(m) \,
%   \sum_{\substack{c>0
%       \\
%       \gamma\in \Gamma_\infty \backslash \Gamma(1) / \Gamma_\infty
%     }}
%   \left[
%     \chi \left(
%       \gamma^{-1}
%     \right)
%   \right]_{\ell_1, \ell_2} \,
%   \frac{ \sqrt{8} \, \pi}{\sqrt{\I \, P} } \, 
%   \sqrt{\ell_2^{~2} - 4 \, P \, m}
%   \\
%   \times 
%   \frac{1}{c^{\frac{3}{2}}} \,
%   \E^{-2 \, \pi \, \I \,
%     \left(
%       \frac{\ell_2^{~2}}{4 \, P} - m
%     \right) \,
%     \frac{a}{c}
%   }
%   \\
  +
  \sum_{\substack{
      n \in \mathbb{Z}
      \\
      n<\frac{a_1^{~2}}{4(k+1)}}}
  q^{n - \frac{a_1^{~2}}{4(k+1)}}
  \sum_{a_2 =1}^{k}
  \sum_{0 \leq m < \frac{a_2^{~2}}{4(k+1)}} p_k^{(a_2)}(m)
  \sum_{\substack{c>0
      \\
      \gamma\in \Gamma_\infty \backslash \Gamma(1) / \Gamma_\infty
    }}
  \left[
    \chi \left(
      \gamma^{-1}
    \right)
  \right]_{a_1, a_2} \,
  \frac{2 \, \pi}{\sqrt{\I}}
  \,
  \left[
    \frac{a_2^{~2} - 4(k+1)m}{a_1^{~2} - 4(k+1)n}
  \right]^{\frac{1}{4}} 
  % \left[
  %   \frac{\Delta(\ell_2)}{ \frac{\ell_1^2}{4 \, P} - n}
  % \right]^{\frac{1}{4}} 
  \\
  \times
  \left[
    1 - E
    \left(
      \sqrt{4 \, \left( \frac{a_1^{~2}}{4(k+1)} -n \right) \,
        \Im(\tau)
      }
    \right)
  \right] 
  \\
  \times
  \frac{1}{c} \, 
  J_{\frac{1}{2}}
  \left(
    \frac{4 \, \pi}{c} \, \sqrt{ \left(
         \frac{a_1^{~2}}{4(k+1)} -n
      \right) \,
%      \Delta(\ell_2)
      \left(
        \frac{a_2^{~2}}{4(k+1)}
        - m
      \right)
    }
  \right) \, 
  \E^{-2  \, \pi \,  \I \,
    \left(
      \frac{a_2^{~2}}{4(k+1)}
      - m
    \right) \,
%    \Delta(\ell_2)
    \frac{a}{c} + 2  \, \pi \, \I \,
    \left(
      n- \frac{a_1^{~2}}{4(k+1)}
    \right) \, 
    \frac{d}{c}
  } .
  \label{Fourier_Poincare_general}
\end{multline}
% Here $\delta_{x\in \mathbb{Z}}$ is $1$ when $x\in \mathbb{Z}$, or
% it vanishes otherwise.

Convergence of this type of series is a delicate
problem~\cite{BrinKOno08a}.
In their work
on the Andrews--Dragonette formula,
Bringmann and Ono proved convergence of such Poincar{\'e} series by
making use of properties of Kloosterman sums and Sali{\'e}
sums~\citep[Section 4]{BrinKOno06a}.
Their proof relies on  the fact that their multiplier system is
parameterized by use of binary quadratic form.
Due to the explicit form of our multiplier system~\eqref{define_chi_a},
their method could be applicable to our
case~\eqref{Fourier_Poincare_general}.
We would like to establish the convergence of the
series~\eqref{Fourier_Poincare_general} mathematically in a future
publication.
We provide a strong evidence for the convergence numerically in
Section~\ref{sec:decomposition}.

Due to Bruinier and Funke~\citep[Proposition 3.2]{BruinFunke04a},
$
\sqrt{\Im \tau} \,
\overline{
  \frac{\partial}{\partial \overline{\tau}}
  P_k^{(a)}(\tau)
}$
has the same modular transformation properties
with $\Psi_k^{(a)}(\tau)$, and
the degrees  of their principal parts coincide.
We have also seen that
the completion $\widehat{\Sigma}_k^{(a)}(\tau)$
fulfills~\eqref{differential_H_Psi}.
We  thus conclude that the Poincar{\'e}--Maass series
$P_k^{(a)}(\tau)$
will coincide with
$\widehat{\Sigma}_k^{(a)}(\tau)$
up to theta functions
when
the polar part~\eqref{def_polar}
is taken from   the Fourier coefficients of
${\Sigma}_k^{(a)}(\tau)$,\footnote{
We would like to thank the referee for pointing out the possible existence of theta function.}
\begin{equation}
  \label{Sigma_and_P}
  \widehat{\Sigma}_k^{(a)}(\tau)
  =
  P_k^{(a)}(\tau) +
  \Theta_k^{(a)}(\tau).
\end{equation}
Here
$\Theta_k^{(a)}\left( 4 \, (k+1) \,\tau\right)$ 
is the theta function on $\Gamma_0\left( 16 \, (k+1)^2 \right)$ with
weight-$1/2$
due to Serre--Stark theorem~\cite{SerreStark77a,KOno08a}.
In the case of $k$ such that
$16 \, (k+1)^2$ is not divisible by $64 \,p^2$ where $p$ is an odd
prime,
or by $4 \, p^2 \, (p^\prime)^2$ with
distinct odd primes $p$ and $p^\prime$,
the theta function $\Theta_k^{(a)}(\tau)$
vanishes.

By dropping the $\bar{\tau}$-dependent parts from the above
formula~\eqref{Fourier_Poincare_general}
we obtain the  holomorphic ($\bar{\tau}$-independent) part which reads
as
\begin{align}
  \Sigma^{(a_1)}_k(\tau) 
  -
  \Theta_k^{(a_1)}(\tau)
  & =
  \left.
    P_k^{(a_1)}(\tau)
  \right|_{\text{holomorphic}}
  \nonumber
  \\
  &  =
  q^{-\frac{a_1^{~2}}{4(k+1)}} \,
  \sum_{ n=0}^\infty
  p_k^{(a_1)}(n) \, q^n 
  .
  % \left(
  %   1 + \sum_{\substack{
  %       n \in \mathbb{Z} \\
  %       n> \frac{\ell_1^{~2}}{4 \, P}
  %     }}
  %   a_P^{(\ell_1)}(n) \, q^n 
  % \right) ,
\end{align}
Since the Fourier coefficients of the weight-1/2 theta function $\Theta_k^{(a_1)}(\tau)$ are constant and do not grow, 
Fourier coefficients of $\Sigma_k^{(a_1)}(\tau)$
are dominated by those of $P_k^{(a_1)}(\tau)$
and each coefficient $p_k^{(a_1)}(n)$
for $n \geq  \frac{a_1^{~2}}{4(k+1)}$ 
is written in terms of
the coefficients of the polar part as
%$q^{n - \frac{\ell_1^2}{4 P}}$ with $n > 0$
\begin{eqnarray}
  \label{p_sum_A}
 p_k^{(a_1)}(n)
  =
  \sum_{a_2=1}^{k}
%  \sum_{0 \leq m < \frac{\ell_2^{~2}}{4 \, P}} 
  \sum_{0 \leq m <
    \frac{a_2^{~2}}{4(k+1)}
  } 
  p_k^{(a_2)}(m)
  \,
  A_k^{(a_2,m,a_1)}(n) ,
\end{eqnarray}
\begin{multline}
  \label{A_exact}
  A_k^{(a_2,m,a_1)}(n)
  =
  \sum_{c=1}^\infty \sum_{\substack{d \mod c\\
      (c,d)=1
    }}
  \sum_{\substack{
      j=0
      \\
     \hskip2mm  j \equiv a_1 \hskip-1mm \mod 2(k+1)
    }}^{2(k+1)c - 1}
  \left(
    \frac{a_2^{~2}- 4(k+1)m}{
      4(k+1)n - a_1^{~2}
    }
  \right)^{\frac{1}{4}} \,
  % \left(
  %   \frac{\Delta(\ell_2)}{
  %     n - \frac{\ell_1^2}{4 \, P}
  %   }
  % \right)^{\frac{1}{4}} \,
  \sqrt{\frac{2 }{ k+1}} \,
  \frac{\pi \, \I}{c^{\frac{3}{2}}}
  \\
  \times
  \E^{2 \, \pi \, \I \, m \, \frac{a}{c} +
    2 \, \pi \, \I \, \left(
      n - \frac{a_1^{~2} - j^2}{4(k+1)} 
    \right) \,
    \frac{d}{c}
  } \,
  \left(
    \E^{ \frac{j  \, a_2}{(k+1)c} \,  \pi \, \I}
    -
    \E^{- \frac{j \,  a_2}{(k+1)c} \,  \pi \, \I}
  \right)
  \\
  \times
  I_{\frac{1}{2}}
  \left(
    \frac{4 \, \pi}{c} \,
    \sqrt{
      \left( n - \frac{a_1^{~2}}{4(k+1)}\right) \,
%      \Delta(\ell_2)
      \left( \frac{a_2^{~2}}{4(k+1)} - m \right)
    }
  \right) .
\end{multline}
Here $a$ is $d^{-1} \mod c$, \emph{i.e.}, $a \, d =1 \mod c$.

The dominant term
of $A_k^{(a_2, m, a_1)}(n)$ comes from a contribution of $c=1$ in
the above infinite series, and we obtain
\begin{multline}
  \label{A_dominating}
  A_k^{(a_2, m, a_1)}(n)
  \approx
%  \sim
  - \pi \, \sqrt{\frac{8}{k+1}} \,
  \sin \left( \frac{a_1 \, a_2}{k+1} \, \pi \right)
  \\
  \times
  \left(
    \frac{a_2^{~2} - 4 \, (k+1) \, m}{
      4 \, (k+1) \, n - a_1^{~2}
    }
  \right)^{\frac{1}{4}} \,
  I_{\frac{1}{2}}
  \left(
    \frac{\pi}{k+1} \,
    \sqrt{
      \left(
        4 \, (k+1) \, n - a_1^{~2}
      \right) \,
      \left(
        a_2^{~2} - 4 \, (k+1) \, m
      \right)
    }
  \right).
\end{multline}

In Refs.~\citenum{DijMalMooVer00a,MansMoor07a,MaloWitt07a}
an expansion of a form similar to~\eqref{p_sum_A}  has been developed 
in the case of holomorphic Jacobi forms (with non-positive weights) 
using the circle method, and the authors discussed the interpretation
of the expansion as a path-integral over  3-dimensional manifolds
related to the BTZ black hole by space-time modular transformations.

%To conclude 
%the Fourier coefficient is
%given in terms of  the coefficients
%$p_k^{(a)}(m)$
%of the principal part
%$0 \leq m < \frac{a^2}{4(k+1)}$,
%which,
%in the case of lower levels,
%corresponds to the massive representations at the unitarity
%bounds~\eqref{recursion_massless_R_tilde}.

%%%%
%%%%%%%%%%%%%
\section{Character Decomposition of Elliptic Genera}
\label{sec:decomposition}
\subsection{Asymptotic Behavior of the Number of Non-BPS Representations}

In our previous paper~\cite{EguchiHikami08a}
we described the general structure of  the elliptic genus
$Z_{X_k}(z;\tau)$
for arbitrary hyperK{\"a}hler
manifold $X_k$
with complex dimension $2\, k$.
Namely  we have shown that it is written as 
\begin{equation}
  \label{elliptic_genus_general}
  Z_{X_k}(z;\tau) =
  Z_{X_k^{(1)}}(z;\tau)
  +
  \sum_{a=2}^{d_k} n_a \, Z_{X_k^{(a)}}(z;\tau),
\end{equation}
where $Z_{X_k^{(a)}}(z;\tau)$
denote symmetric polynomials of the ratios of Jacobi theta
functions,
$\left(
  \frac{\theta_{10}(z;\tau)}{\theta_{10}(0;\tau)}
\right)^2$,
$\left(
  \frac{\theta_{00}(z;\tau)}{\theta_{00}(0;\tau)}
\right)^2$,
and
$\left(
  \frac{\theta_{01}(z;\tau)}{\theta_{01}(0;\tau)}
\right)^2$ 
of order-$k$.
Each $Z_{X_k^{(a)}}(z;\tau)$ is
a  Jacobi form with weight-$0$ and index-$k$,
and $d_k$ denotes a dimension of the space of these Jacobi forms.
The normalization of $Z_{X_k^{(a)}}$ is fixed so that its
$q$-expansion has integer coefficients~\cite{EguchiHikami08a}.
Amongst others, we have set $Z_{X_k^{(1)}}(z;\tau)$ to be
\begin{equation}
  Z_{X_k^{(1)}}(z;\tau)
  =
  (k+1) \, 2^{2 \,k}
  \,
  \left[
    \left(
      \frac{\theta_{10}(z;\tau)}{\theta_{10}(0;\tau)}
    \right)^{2 \, k}
    +
    \left(
      \frac{\theta_{00}(z;\tau)}{\theta_{00}(0;\tau)}
    \right)^{2 \, k}
    +
    \left(
      \frac{\theta_{01}(z;\tau)}{\theta_{01}(0;\tau)}
    \right)^{2 \, k}
  \right] ,\label{leading}
\end{equation}
where the prefactor is chosen so that the identity representation in the NS
sector has a multiplicity~$1$ in partition function ~\cite{EguchiHikami08a}.
The identity representation comes  only from $Z_{X_k^{(1)}}(z;\tau)$,
so the elliptic genus of $X_k$  can be determined to be of the
form~\eqref{elliptic_genus_general}, (\ref{leading}).

Among  hyperK{\"a}hler manifolds, the Hilbert scheme of points on
the $K3$ surface 
$K3^{[m]}$
has been much studied.
It was proposed
by a method of the second quantized string
that their elliptic genera are obtained
as~\cite{DijMooVerVer97a}
\begin{equation}
  \label{DMVV_K3}
  \sum_{m=0}^\infty p^m \,
  Z_{K3^{[m]}}(z;\tau)
  =
  \prod_{n=1}^\infty
  \prod_{m=0}^\infty \prod_{\ell \in \mathbb{Z}}
  \frac{1}{
    \left(
      1 - p^n \, q^m \, \zeta^\ell
    \right)^{c(n \, m , \ell)}
  } ,
\end{equation}
where $c(n,\ell)$ is the Fourier coefficients of the elliptic genus
for the $K3$ surface,
\begin{equation*}
  2 \, \phi_{0,1}(z;\tau)
  =
  \sum_{n=0}^\infty \sum_{\ell \in \mathbb{Z}}
  c(n,\ell) \, q^n \, \zeta^\ell .
\end{equation*}
This generating function~\eqref{DMVV_K3} 
is a generalization of the identity
for the Euler characteristics~\cite{Gotts90a}.

It is known
(see, \emph{e.g.}, Ref.~\citenum{EWitt87a})
that the classical topological invariants of $X$,
such as the
Euler character, the Hirzebruch signature, and the
$\widehat{A}$-genus,
are respectively given by
\begin{equation}
  \begin{gathered}
    Z_{X_k}(z=0;\tau) = \chi_{X_k} ,
    \\[2mm]
    Z_{X_k}
    \left( z=\frac{1}{2} ;\tau \right)
    =
    \sigma_{X_k} + \cdots,
    \\[2mm]
    (-1)^k \, q^{\frac{k}{2}} \, Z_{X_k} \left(
      z=\frac{1+\tau}{2};\tau
    \right)=
    \widehat{A}_{X_k}+ \cdots .
  \end{gathered}
\end{equation}

It is easy to see that the only contribution to the $\widehat{A}$-genus
comes from the leading term 
$Z_{X_k^{(1)}}$ in~\eqref{elliptic_genus_general} 
and we easily find 
\begin{equation}
  \widehat{A}_{X_k}=k+1 
\end{equation}
for any hyperK\"ahler manifolds in $2k$ complex dimensions.
It turned out that this result has been known in the mathematical literature ~\cite{HitcSawo01a}.

Now we present an  estimate on the asymptotic behavior of the number
of non-BPS representations in general hyperK{\"a}hler manifold $X_k$.
We first decompose the elliptic genus~\eqref{elliptic_genus_general}
into a sum over characters 
\begin{equation}
  Z_{X_k}(z;\tau)
  =
  \chi_{X_k} \cdot
  C_{k}(z;\tau)
  -
  \sum_{a=1}^k
  \Sigma_{X_k}^{(a)}(\tau) \,
  B_{k}^{(a)}(z;\tau)  ,
  \label{character-decompo}
\end{equation}
where $\Sigma_{X_k}^{(a)}(\tau)$ has an expansion of the form
\begin{equation}
  \Sigma_{X_k}^{(a)}(\tau)
  =
  \sum_{n=0}^{\infty} p_k^{(a)}(n) \,
  q^{n-{a^2\over 4(k+1)}}.
\label{Sigma}
\end{equation}
Due to discussions in Section~\ref{sec:integrality}, we have
$
p_k^{(a)}(n)\in \mathbb{Z}$.
Here $n=0$ corresponds to the unitarity boundary.
As we know, massive
representations at the unitarity boundary are decomposed into massless
representations. 
Thus the $n=0$ pieces in~\eqref{Sigma} are  absorbed
into the first part of~\eqref{character-decompo} and then the sum over
$n$ in~\eqref{Sigma} runs from $n=1$ to $\infty$.

On the other hand, if we look at the expressions~\eqref{p_sum_A}
and~\eqref{A_dominating}, we find 
\begin{multline*}
  p_k^{(a_1)}(n)
  \approx 
  - \pi \, \sqrt{\frac{8}{k+1}} \,
  \sum_{a_2=1}^k
  \sin \left(
    \frac{a_1 \, a_2}{k+1} \, \pi
  \right)
  \sum_{0\leq m< {a_2^{~2}\over  4(k+1)}}
  p_k^{(a_2)}(m) 
  \\
  \times
  \left(
    \frac{
      a_2^{~2}-4 \, (k+1) \, m
    }{
      4 \, (k+1) \, n - a_1^{~2}
    }
  \right)^{1/4} \,
  I_{{1\over 2}}
  \left(
    {\pi \over    k+1} \,
    \sqrt{
      \left(
        4 \, (k+1) \, n-a_1^{~2}
      \right) \,
      \left(
        a_2^{~2}-4 \, (k+1) \,m
      \right)}
  \right).
\end{multline*}
Since the Bessel function $I_{{1\over 2}}(x)$ is
$\sqrt{{2\over \pi  \, x}} \, \sinh(x)$, 
the 
dominant contribution to the asymptotic behavior of the coefficients
$p_k^{(a_1)}(n)$ comes from the largest value of $a_2(=k)$ and the
smallest value of $m(=0)$ in the polar part of ~\eqref{Sigma}. 
It is fairly easy to see that the term with maximal isospin
${a_2/2}={k/2}$ at the 
unitarity boundary $m=0$,  \emph{i.e.}, 
$p_k^{(k)}(0)$ comes only from the leading term~\eqref{leading} of the
elliptic genus and equals to $p_k^{(k)}(0)=k+1$.
We get
\begin{equation}
  p_k^{(a)}(n)
  \sim
  (-1)^a \, \pi \,
  \frac{
    \sqrt{8 \, k \, (k+1)}}{
    \left(
      4 \, (k+1) \, n - a^2
    \right)^{1/4}
  }  \,
  \sin \left( \frac{a}{k+1} \, \pi \right) \,
  I_{\frac{1}{2}}
  \left(
    \frac{k \, \pi}{k+1} \,
    \sqrt{
      4 \, (k+1) \, n - a^2
    }
  \right) .
\end{equation}
Thus quite generally,
independent of the values of $n_a$ in~\eqref{elliptic_genus_general},
we obtain the asymptotic estimate 
\begin{equation}
  \left|
    p_k^{(a)}(n)
  \right|
%  \approx
  \sim
  \exp
  \left(
    2 \, \pi \, \sqrt{{k^2\over k+1} \, n
      -
      \left({k\over k+1}\cdot {a\over 2}\right)^2}
  \right).
\end{equation}

The level-1 case, $k=1$ and $a=1$,
is a result in our previous
paper~\cite{EguchiHikami09a}.

%%%%%%%%%%%%%%%%%%%%
\subsection{Examples}
\subsubsection{Level-$2$}

We present in some detail the results for the hyperK{\"a}hler manifolds $X_2(n)$ of
complex dimension $4$.
The elliptic genus of $X_2(n)$ is the Jacobi form with weight-$0$ and
index-$2$,
and it is a linear combination of
$
    \left[ \phi_{0,1} \right]^2 
$ and
$
    \left[ \phi_{-2,1} \right]^2 \, E_4
$ in~\eqref{Eichler_Zagier_base}.
In our previous paper~\cite{EguchiHikami08a}, we set bases of the
Jacobi forms to be
\begin{align}
  Z_{X_2^{(1)}}(z;\tau)
  & =
  48 \,
    \left[
    \left(
      \frac{\theta_{10}(z;\tau)}{\theta_{10}(0;\tau)}
    \right)^4
    +
    \left(
      \frac{\theta_{00}(z;\tau)}{\theta_{00}(0;\tau)}
    \right)^4
    +
    \left(
      \frac{\theta_{01}(z;\tau)}{\theta_{01}(0;\tau)}
    \right)^4
  \right] ,
  \\[2mm]
  Z_{X_2^{(2)}}(z;\tau)
  & =
  2 \,
  \Biggl[
    \left(
      \frac{\theta_{10}(z;\tau)}{\theta_{10}(0;\tau)} \cdot
      \frac{\theta_{00}(z;\tau)}{\theta_{00}(0;\tau)}
    \right)^2
    +
    \left(
      \frac{\theta_{00}(z;\tau)}{\theta_{00}(0;\tau)} \cdot
      \frac{\theta_{01}(z;\tau)}{\theta_{01}(0;\tau)}
    \right)^2
    \nonumber
    \\
    & \qquad \qquad \qquad
    +
    \left(
      \frac{\theta_{01}(z;\tau)}{\theta_{01}(0;\tau)} \cdot
      \frac{\theta_{10}(z;\tau)}{\theta_{10}(0;\tau)}
    \right)^2
  \Biggr] .
\end{align}
They are identified with
\begin{equation}
  \begin{pmatrix}
    \left[ \phi_{0,1} \right]^2
    \\[1mm]
    \left[ \phi_{-2,1} \right]^2 \, E_4
  \end{pmatrix} 
  =
  \begin{pmatrix}
    1 & 16
    \\
    1 & - 8
  \end{pmatrix} \,
  \begin{pmatrix}
    \frac{1}{3} \, Z_{X_2^{(1)}}(z;\tau)
    \\[1mm]
    Z_{X_2^{(2)}}(z;\tau)
  \end{pmatrix} ,
\end{equation}
% \begin{equation}
%   \begin{pmatrix}
%     Z_{X_2^{(1)}}(z;\tau)
%     \\[1mm]
%     Z_{X_2^{(2)}}(z;\tau)
%   \end{pmatrix}
%   =
%   \begin{pmatrix}
%     1 & 2
%     \\
%     \frac{1}{24} & - \frac{1}{24}
%   \end{pmatrix} \,
%   \begin{pmatrix}
%     \left[ \phi_{0,1} \right]^2
%     \\[1mm]
%     \left[ \phi_{-2,1} \right]^2 \, E_4
%   \end{pmatrix} ,
% \end{equation}
and in the notation of Ref.~\citenum{Gritse99a} we have
\begin{equation*}
  Z_{X_2^{(2)}}(z;\tau) 
  = \phi_{0,2}(z;\tau) 
  .
\end{equation*}

The elliptic genus for dimension-$4$ 
manifold $X_2(n)$ is defined  as
\begin{equation*}
  Z_{X_2(n)}(z;\tau)
  =
  Z_{X_2^{(1)}}(z;\tau)
  + n \,
  Z_{X_2^{(2)}}(z;\tau),
\end{equation*}
which gives
\begin{equation}
  \begin{aligned}
    Z_{X_2(n)}(0;\tau)
    & =
    144+ 6 \, n ,
    \\[2mm]
    Z_{X_2(n)}
    \left( \frac{1}{2};\tau \right)
    & =
    \left(
      96 +2 \, n
    \right) + \cdots,
    \\
    Z_{X_2(n)}
    \left( \frac{1+\tau}{2};\tau \right)
    & =
    3 \, q^{-1} + \cdots .
  \end{aligned}
\end{equation}
Especially we have
$K3^{[2]}=X_2(n=30)$,
\begin{equation}
  \begin{aligned}
    \chi_{K3^{[2]}}
    & =
    324 ,
    &
    \sigma_{K3^{[2]}}
    & = 156 
    ,
    &
    \widehat{A}_{K3^{[2]}} 
    & = 3 .
  \end{aligned}
\end{equation}

Using the character decomposition ~\eqref{J_decompose_C_H},
we have~\cite{EguchiHikami08a}
% \begin{equation}
%   \begin{aligned}
%     \ch_{k=2,h=\frac{2}{4},\ell=0}^{\widetilde{R}}(z;\tau)
%     & =
%     \left(
%       \frac{\theta_{10}(z;\tau)}{
%         \theta_{10}(0;\tau)}
%     \right)^4
%     +
%     \sum_{a=1}^2
%     H_3^{(a)}\left(\boldsymbol{w}_{2,0,0};\tau \right) \,
%     B_3^{(a)}(z;\tau)
%     \\
%     & =
%     \left(
%       \frac{\theta_{00}(z;\tau)}{
%         \theta_{00}(0;\tau)}
%     \right)^4
%     +
%     \sum_{a=1}^2
%     H_3^{(a)}\left(\boldsymbol{w}_{0,2,0};\tau \right) \,
%     B_3^{(a)}(z;\tau)
%     \\
%     & =
%     \left(
%       \frac{\theta_{01}(z;\tau)}{
%         \theta_{01}(0;\tau)}
%     \right)^4
%     +
%     \sum_{a=1}^2
%     H_3^{(a)}\left(\boldsymbol{w}_{0,0,2};\tau \right) \,
%     B_3^{(a)}(z;\tau)
%   \end{aligned}
% \end{equation}
% and that
\begin{equation}
  \label{genus_level2}
  \begin{aligned}
    Z_{X_2^{(1)}}(z;\tau)
    & = 
    144 \, \ch_{k=2,h=\frac{2}{4},\ell=0}^{\widetilde{R}}(z;\tau)
    -
    \sum_{a=1}^2
    \Sigma_{(2,0,0)}^{(a)}(\tau) \,
  % \sum_{\boldsymbol{k}:
  %   \text{sym$(2,0,0)$}
  % }
  % 48 \,
  % H_3^{(a)} \left( \boldsymbol{w}_{\boldsymbol{k}};\tau \right) \,
    B_2^{(a)}(z;\tau) ,
    \\[2mm]
    Z_{X_2^{(2)}}(z;\tau)
    & = 
    6 \, \ch_{k=2,h=\frac{2}{4},\ell=0}^{\widetilde{R}}(z;\tau)
    -
    \sum_{a=1}^2
    \Sigma_{(1,1,0)}^{(a)}(\tau) \,
  % \sum_{\boldsymbol{k}:
  %   \text{sym$(1,1,0)$}
  % }
  % 2 \,
  % H_3^{(a)} \left( \boldsymbol{w}_{\boldsymbol{k}};\tau \right) \,
    B_2^{(a)}(z;\tau) .
  \end{aligned}
\end{equation}
Here  the functions $\Sigma^{(*)}_*(\tau)$ are expanded as 
\begin{gather*}
  \begin{aligned}
    % \sum_{\boldsymbol{k}:
    %   \text{sym$(2,0,0)$}
    % }
    % 48 \,
    % \begin{pmatrix}
    %   H_3^{(1)} \left( \boldsymbol{w}_{\boldsymbol{k}};\tau \right)
    %   \\
    %   H_3^{(2)} \left( \boldsymbol{w}_{\boldsymbol{k}};\tau \right)
    % \end{pmatrix}
    \begin{pmatrix}
      {\Sigma}^{(1)}_{(2,0,0)}(\tau)\\
      {\Sigma}^{(2)}_{(2,0,0)}(\tau)
    \end{pmatrix}
    &
    =
    \begin{pmatrix}
      q^{-\frac{1}{12}} \,
      \left[
        18
        - 1872 \, q
        - 26070 \, q^2
        - 213456 \, q^3
        - 1311420 \, q^4
        - \cdots
      \right]
      \\[1mm]
      q^{- \frac{1}{3}} \,
      \left[
        3
        + 510 \, q
        + 12804 \, q^2
        + 126360 \, q^3
        + 841176 \, q^4
        + \cdots
      \right]
    \end{pmatrix} ,
  \end{aligned}
  \\[3mm]
  \begin{aligned}
    % \sum_{\boldsymbol{k}:
    %   \text{sym$(1,1,0)$}
    % }
    % 2 \,
    % \begin{pmatrix}
    %   H_3^{(1)} \left( \boldsymbol{w}_{\boldsymbol{k}};\tau \right)
    %   \\
    %   H_3^{(2)} \left( \boldsymbol{w}_{\boldsymbol{k}};\tau \right) 
    % \end{pmatrix}
    \begin{pmatrix}
      {\Sigma}^{(1)}_{(1,1,0)}(\tau)\\
      {\Sigma}^{(2)}_{(1,1,0)}(\tau) 
    \end{pmatrix} 
    & 
    =
    \begin{pmatrix}
      q^{-\frac{1}{12}} \,
      \left[
        1
        - 16 \, q
        - 55 \, q^2
        - 144 \, q^3
        - 330 \, q^4
        - \cdots
      \right]
      \\[1mm]
      q^{- \frac{1}{3}} \,
      \left[
        - 10 \, q
        - 44 \, q^2
        - 110 \, q^3
        - 280 \, q^4
        - \cdots
      \right]
    \end{pmatrix}
    .
%     \\
%     & 
%     \sim
%     \begin{pmatrix}
%       \displaystyle
%       q^{-\frac{1}{12}} 
%       - 
%       \sum_{n=1}^\infty
%       \frac{
%         \sqrt{2} \, \pi
%       }{
%         \left( 12 \, n - 1 \right)^{\frac{1}{4}}
%       } \,
%       I_{\frac{1}{2}} 
%       \left(
%         \frac{\pi}{3} \, \sqrt{ 12 \, n - 1}
%       \right)
%       \,
%       q^{n-\frac{1}{12}}
%       \\[1mm]
% %
%       \displaystyle
%       -
%       \sum_{n=1}^\infty
%       \frac{
%         \sqrt{2} \, \pi
%       }{
%         \left( 12 \, n - 4 \right)^{\frac{1}{4}}
%       } \,
%       I_{\frac{1}{2}} 
%       \left(
%         \frac{\pi}{3} \, \sqrt{ 12 \, n - 4}
%       \right)
%       \, q^{n-\frac{1}{3}}
%     \end{pmatrix}
%     .
  \end{aligned}
\end{gather*}
The polar parts are
\begin{align*}
  \left.
    \begin{pmatrix}
      {\Sigma}^{(1)}_{(2,0,0)}(\tau)\\
      {\Sigma}^{(2)}_{(2,0,0)}(\tau) 
    \end{pmatrix}
  \right|_{\text{polar}}
  &=
  \begin{pmatrix}
    18 \, q^{-\frac{1}{12}}
    \\
    3 \, q^{-\frac{1}{3}}
  \end{pmatrix}
  ,
  \\[2mm]
  \left.
    \begin{pmatrix}
      {\Sigma}^{(1)}_{(1,1,0)}(\tau)\\
      {\Sigma}^{(2)}_{(1,1,0)}(\tau) 
    \end{pmatrix}
  \right|_{\text{polar}}
  &=
  \begin{pmatrix}
    q^{-\frac{1}{12}} 
    \\
    0
  \end{pmatrix}
  .
\end{align*}
These
are massive characters at the unitarity bound,
which are decomposed into a sum of massless
characters.
Then we obtain
\begin{gather*}
  \begin{aligned}
    Z_{X_2^{(1)}}(z;\tau)
    & =
    111 \, \ch_{2,\frac{2}{4},0}^{\widetilde{R}}(z;\tau)
    -
    12 \, \ch_{2,\frac{2}{4},\frac{1}{2}}^{\widetilde{R}}(z;\tau)
    +
    3 \, \ch_{2,\frac{2}{4},1}^{\widetilde{R}}(z;\tau)
    \\ &
    +
      q^{-\frac{1}{12}} \,
      \left[
        1872 \, q + 26070 \, q^2
        +213456 \, q^3 + 1311420 \, q^4 + \cdots
      \right]B_2^{(1)}(z;\tau)
      \\&
+
      q^{-\frac{1}{3}} \,
      \left[
        -510 \, q - 12804 \, q^2
        -126360 \, q^3 - 841176 \, q^4 - \cdots
      \right]B_2^{(2)}(z;\tau)    
    ,
  \end{aligned}
  \\[2mm]
  \begin{aligned}
    Z_{X_2^{(2)}}(z;\tau)
    & =
    4 \, \ch_{2,\frac{2}{4},0}^{\widetilde{R}}(z;\tau)
    -
    \ch_{2,\frac{2}{4},\frac{1}{2}}^{\widetilde{R}}(z;\tau)
    \\ & 
    +q^{-\frac{1}{12}} \,
      \left[
        16 \, q + 55 \, q^2
        + 144 \, q^3 + 330 \, q^4 + \cdots
      \right]B_2^{(1)}(z;\tau)
      \\&
+  q^{-\frac{1}{3}} \,
      \left[
        10 \, q +44 \, q^2
        + 110 \, q^3 +280 \, q^4 + \cdots
      \right]B_2^{(2)}(z;\tau)
    .
  \end{aligned}
\end{gather*}

\begin{figure}
  \centering
  \includegraphics[scale=1.0]{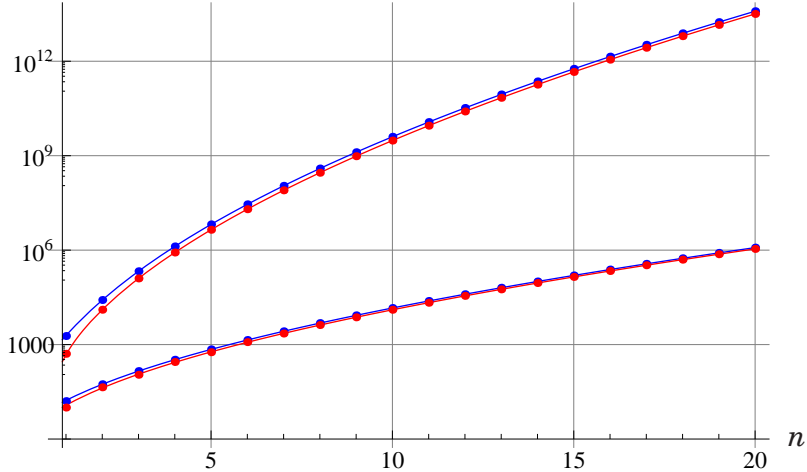}
  \caption{Absolute values of the Fourier coefficients of level-$2$.
    Blue dots and curves  denote respectively
    exact and asymptotic values 
    of 
    $\Sigma_{(2,0,0)}^{(1)}$
    and $\Sigma_{(1,1,0)}^{(1)}$.
    Red dots and curves are for
    $\Sigma_{(2,0,0)}^{(2)}$
    and $\Sigma_{(1,1,0)}^{(2)}$.
  }
  \label{fig:level2}
\end{figure}

In Fig.~\ref{fig:level2} 
we have  plotted
both the exact values (obtained from using \eqref{h_Fourier_J}) and
the prediction of the asymptotic formula~\eqref{A_dominating} for  
(the absolute values of) the expansion
coefficients of $\Sigma^{(a)}_{(2,0,0)}$ and $\Sigma^{(a)}_{(1,1,0)}$.
The theta function in~\eqref{Sigma_and_P} vanishes in this case.

In order to check the convergence of our results  
we also present some numerical data in the table:
here the results obtained by truncating the infinite sum over
$c$~\eqref{A_exact} 
at $c=1$,
$5$ and $50$ are presented. We see a very fast convergence.
\begin{itemize}
\item
  the Fourier
  coefficients,
  $\Coeff_{q^{n-\frac{a^2}{12}}} \left[
    {\Sigma}_{(2,0,0)}^{(a)}(\tau)
  \right]$,
\begin{equation*}
%  \small
  \scriptsize
  \begin{array}{cc||r|rrr}
    n & a & \text{exact} & \sum_{c=1}^1 & \sum_{c=1}^5 & \sum_{c=1}^{50}
    \\
    \hline \hline
    2 &1& -26070 & -25934.120 & -26058.697 & -26072.610
    \\
    &2& 12804 & 12827.954 & 12822.271 &  12803.513
    \\
    \hline
    4 &1
    & -1311420 
    & -1310418.583
    & -1311430.279
    & -1311415.819
    \\
    & 2
    & 841176 
    & 841279.585
    & 841175.261
    & 841178.319
    \\
    \hline
    10 & 1
    & -3984136794 
    & -3984092994.253
    & -3984136778.572
    & -3984136798.536
    \\
    & 2
    & 3019548204
    & 3019548311.055
    & 3019548172.087
    & 3019548207.266
    \\
    \hline
    20 &1
    & -38753796654252
    & -38753793062898.157
    & -38753796654206.172
    & -38753796654250.587
    \\
    & 2
    & 31807078711584 
    & 31807078712552.794
    & 31807078711534.247
    & 31807078711584.222
    \\
    \hline
  \end{array}
\end{equation*}

\item the Fourier
  coefficients,
  $\Coeff_{q^{n-\frac{a^2}{12}}} 
  \left[
    {\Sigma}_{(1,1,0)}^{(a)}(\tau)
  \right]$,
\begin{equation*}
%  \small
  \scriptsize
  \begin{array}{cc||r|rrr}
    n &a& \text{exact} & \sum_{c=1}^1 & \sum_{c=1}^5 & \sum_{c=1}^{50}
    \\
    \hline \hline
    2 & 1
    & -55 & -54.800 & -54.533 & -55.128
    \\
    & 2
    & -44 & -41.870 & -43.018 & -44.040
    \\
    \hline
    4 & 1
    &  -330 & -331.443 & -330.415 & -329.790
    \\
    &2
    & -280 & -271.384 & -280.221 & -279.897
    \\
    \hline
    10 & 1
    & -14509
    & -14520.562
    & -14507.586
    & -14509.198
    \\
    & 2
    & -12772
    & -12723.091 
    & -12773.505
    & -12771.797
    \\
    \hline
    20 &1
    & -1203058
    & -1203032.050
    & -1203057.702
    & -1203057.897
    \\
    & 2
    & -1093664
    & -1093336.160
    & -1093664.465
    & -1093664.021
    \\
    \hline
  \end{array}
\end{equation*}
\end{itemize}

%%%%%%%
We shall comment on  a relationship between $Z_{X_2^{(1)}}$ and
$Z_{X_2^{(2)}}$, and show that the Poincar{\'e}--Maass
series~\eqref{Fourier_Poincare_general} is merely a holomorphic Jacobi
form when the polar parts are suitably chosen.
The Riemann addition formulae for the Jacobi theta series
(see, \emph{e.g.}, Ref.~\citenum{Mumf83})
read as
\begin{equation}
  \begin{aligned}[b]
    \left(
      \frac{\theta_{01}(z;\tau)}{
        \theta_{01}(0;\tau)}
    \right)^2
    & =
    \left(
      \frac{\theta_{10}(z;\tau)}{
        \theta_{10}(0;\tau)}
    \right)^2
    +
    \frac{1}{4} \,
    \left(
      \frac{\theta_{00}(0;\tau)}{\eta(\tau)}
    \right)^4 \,
    \left(
      \frac{\theta_{11}(z;\tau)}{\eta(\tau)}
    \right)^2
    \\
    & =
    \left(
      \frac{\theta_{00}(z;\tau)}{
        \theta_{00}(0;\tau)}
    \right)^2
    +
    \frac{1}{4} \,
    \left(
      \frac{\theta_{10}(0;\tau)}{\eta(\tau)}
    \right)^4 \,
    \left(
      \frac{\theta_{11}(z;\tau)}{\eta(\tau)}
    \right)^2 ,
  \end{aligned}
\end{equation}
which shows
\begin{eqnarray}
  Z_{X_2^{(1)}}(z;\tau)
  -
  24 \, Z_{X_2^{(2)}}(z;\tau)
  & =&
  3 \,
  \frac{E_4(\tau)}{
    \left[ \eta(\tau) \right]^{10}
  } \,
  \left[
    \Psi_2^{(2)}(\tau) \, B_2^{(1)}(z;\tau)
    -
    \Psi_2^{(1)}(\tau) \, B_2^{(2)}(z;\tau)
  \right]
  \nonumber
  \\
  \label{identity_square_phi2}
  & =&
  3 \, \left[
    \phi_{-2,1}(z;\tau)
  \right]^2 \, E_4(\tau) .
\end{eqnarray}
where $\Psi^{(a)}_k(\tau)$'s are defined in (\ref{def_Psi}).

(\ref{identity_square_phi2}) shows that $\boldsymbol{\Phi}_2(\tau)$
defined by
\begin{align}
  {\boldsymbol{\Phi}}_2(\tau)
  & =
  \frac{E_4(\tau)}{
    \left[ \eta(\tau) \right]^{10}
  } \,
  \begin{pmatrix}
    \Psi_2^{(2)}(\tau)
    \\[1mm]
    - \Psi_2^{(1)}(\tau)
  \end{pmatrix}
  =
  \frac{E_4(\tau)}{
    \left[ \eta(\tau) \right]^{10}
  } \,
  \begin{pmatrix}
    2 \,
    \frac{\left[ \eta(\tau) \, \eta(4 \, \tau) \right]^2}{
      \eta(2 \, \tau)}
    \\[1mm]
    -
    \frac{\left[  \eta(2 \, \tau) \right]^5}{
      \left[ \eta(4 \, \tau) \right]^2}
  \end{pmatrix}
  \\
  \nonumber
  & =
  \begin{pmatrix}
    q^{-\frac{1}{12}} \,
    \left[
      2 + 496 \, q
      + 8250 \, q^2
      + 70000 \, q^3
      + 434500 \, q^4
      + 2184512 \, q^5
      + \cdots
    \right]
    \\[1mm]
    q^{-\frac{1}{3}} \,
    \left[
      -1
      - 250 \, q
      - 4620 \, q^2
      - 43000 \, q^3
      - 282632 \, q^4
      - 1484252 \, q^5
      - \cdots
    \right]
  \end{pmatrix}
  ,
\end{align}
is a holomorphic vector-valued modular form 
which transforms as~\eqref{general_transform_Poincare} with $k=2$.
% We note that
% \begin{equation}
%   \label{identity_square_phi2}
%   \left[
%     \phi_{-2,1}
%   \right]^2
%   =
%   \frac{1}{
%     \left[
%       \eta(\tau)
%     \right]^{10}
%   } \,
%   \left[
%     \Psi_3^{(2)}(\tau) \, B_3^{(1)}(z;\tau)
%     -
%     \Psi_3^{(1)}(\tau) \, B_3^{(2)}(z;\tau)
%   \right]
% \end{equation}
%%%%%
% As was discussed in the previous section,
% one of
% solutions of  \eqref{Laplacian_Poincare}
% satisfying~\eqref{ST_transform_Poincare} for $P=3$
% are given by the Poincar{\'e}--Maass series
% $
% \boldsymbol{P}_3(\tau)=
% \begin{pmatrix}
%   P_3^{(1)}(\tau) \\
%   P_3^{(2)}(\tau)
% \end{pmatrix}
% $ defined in~\eqref{define_Poincare-Maass}.
%which  follows from~\eqref{h_Fourier_J}
Existence of this form follows from the fact
that
one of two Jacobi forms with index-2,
$\left[ \phi_{-2,1} \right]^2 \, E_4$,
vanishes at $z=0$.
%The identity~\eqref{identity_square_phi2} is a consequence
%of~\eqref{h_Fourier_J}
In fact when we 
substitute
$\left[ \phi_{-2,1} \right]^2 \, E_4$
for
$\widehat{J}_k(z;\tau)$ in~\eqref{h_Fourier_J}
\begin{equation*}
  \boldsymbol{\Phi}_2(\tau)
  =
  - \I \,
  \left[ \eta(\tau) \right]^3 \, E_4(\tau) \,
  \int_{z_0}^{z_0+1}
  \frac{
    \theta_{11}(2 \, z; \tau)
  }{
    \left[
      \theta_{11}(z;\tau)
    \right]^2
  } \,
  \left[ \phi_{-2,1}(z; \tau) \right]^2 \,
  \begin{pmatrix}
    q^{-\frac{1}{12}} \, \E^{-2 \, \pi \, \I \, z}
    \\[1mm]
    q^{- \frac{1}{3}} \, \E^{-4 \, \pi \, \I \, z}
  \end{pmatrix} \,
  \mathrm{d} z,
\end{equation*}
the integrand is non-singular and is well-defined because of the zero of
$\left[ \phi_{-2,1} \right]^2$. 

From~\eqref{genus_level2},
we obtain
\begin{equation*}
  \boldsymbol{\Phi}_2(\tau)
  =
  -{1\over 3} \,
  \begin{pmatrix}
    \Sigma_{(2,0,0)}^{(1)}(\tau)
    \\
    \Sigma_{(2,0,0)}^{(2)}(\tau)
  \end{pmatrix}
  +
  8 \,
  \begin{pmatrix}
    \Sigma_{(1,1,0)}^{(1)}(\tau)
    \\
    \Sigma_{(1,1,0)}^{(2)}(\tau)
  \end{pmatrix}
  .
%   -16 \,
%   \left[
%     \boldsymbol{\Sigma}_{(2,0,0)}(\tau)
%     -
%     \boldsymbol{\Sigma}_{(1,1,0)}(\tau)
%   \right] ,
\end{equation*}
In these combinations $\Sigma$ functions acquire good modular
transformation properties.
\subsubsection{Level-$3$}
We have three Jacobi forms with weight-$0$ and index-$3$.
Each Jacobi form $Z_{X_3^{(a)}}$ is defined and
decomposed as follows;
\begin{gather*}
  \begin{aligned}
    \frac{1}{4}  \, Z_{X_3^{(1)}}(z;\tau)
    & =64 \,
    \left[
      \left(
        \frac{\theta_{10}(z;\tau)}{
          \theta_{10}(0;\tau)}
      \right)^6
      + \text{others}
%      X^3 + Y^3 + Z^3
    \right]
    \\
    & =
    142 \ch_{3,\frac{3}{4},0}^{\widetilde{R}}(z;\tau)
    - 14 \, \ch_{3,\frac{3}{4},\frac{1}{2}}^{\widetilde{R}}(z;\tau)
    + 6  \,  \ch_{3,\frac{3}{4},1}^{\widetilde{R}}(z;\tau)
    -    \ch_{3,\frac{3}{4},\frac{3}{2}}^{\widetilde{R}}(z;\tau)
    \\
    & \qquad
    +
    q^{-\frac{1}{16}} \,
    \left[
      5306 \, q + 145410 \, q^2 + 2248049 \, q^3
      +
      \cdots
    \right]  \, B_3^{(1)}(z;\tau)
    \\
    &  \qquad 
    +
    q^{-\frac{1}{4}} \,
    \left[
      -1856 \, q - 97368 \, q^2  - 1848000 \, q^3
      + \cdots
    \right] \, B_3^{(2)}(z;\tau)
    \\
    & \qquad 
    +
    q^{-\frac{9}{16}} \,
    \left[
      -21 \, q + 17927 \, q^2 + 510797 \, q^3
      + \cdots
    \right] \, B_3^{(3)}(z;\tau)
    ,
  \end{aligned}
  \\
  \begin{aligned}
    Z_{X_3^{(2)}}(z;\tau)
    & =
    8 \, \left[
      \left(
        \frac{\theta_{10}(z;\tau)}{
          \theta_{10}(0;\tau)}
      \right)^4 \,
      \left(
        \frac{\theta_{00}(z;\tau)}{
          \theta_{00}(0;\tau)}
      \right)^2
%      X^2 \, Y
      + \text{others}
    \right]
    \\
    & =
    29 \ch_{3,\frac{3}{4},0}^{\widetilde{R}}(z;\tau)
    - 8 \, \ch_{3,\frac{3}{4},\frac{1}{2}}^{\widetilde{R}}(z;\tau)
    + \ch_{3,\frac{3}{4},1}^{\widetilde{R}}(z;\tau)
    \\
    & \qquad
    +
    q^{-\frac{1}{16}} \,
    \left[
      294 \, q + 2466 \, q^2 + 14302 \, q^3
      + \cdots
    \right] \, B_3^{(1)}(z;\tau)
    \\
    & \qquad 
    +
    q^{-\frac{1}{4}} \,
    \left[
      72 \, q + 261 \, q^2 + 504 \, q^3 +
      \cdots
    \right] \, B_3^{(2)}(z;\tau)
    \\
    & \qquad 
    + 
    q^{-\frac{9}{16}} \,
    \left[
      -18 \, q - 644 \, q^2 -5544 \, q^3
      - \cdots
    \right] \, B_3^{(3)}(z;\tau)
    ,
  \end{aligned}
  \\[2mm]
%  \displaybreak[0]
%
  \begin{aligned}
    Z_{X_3^{(3)}}(z;\tau)
    & =
    4 \,
    \left(
      \frac{\theta_{10}(z;\tau)}{
        \theta_{10}(0;\tau)}
    \right)^2
    \,
    \left(
      \frac{\theta_{00}(z;\tau)}{
        \theta_{00}(0;\tau)}
    \right)^2
    \,
    \left(
      \frac{\theta_{01}(z;\tau)}{
        \theta_{01}(0;\tau)}
    \right)^2
%    4 \, X \, Y \, Z
    \\
    & =
    2 \ch_{3, \frac{3}{4},0}^{\widetilde{R}}(z;\tau)
    - \ch_{3, \frac{3}{4}, \frac{1}{2}}^{\widetilde{R}}(z;\tau)
    \\
    & \qquad
    +
    q^{-\frac{1}{16}} \,
    \left[
      7 \, q + 21 \, q^2 + 43 \, q^3 + 94 \, q^4
      + \cdots
    \right]  \, B_3^{(1)}(z;\tau)
    \\
    & \qquad 
    +
    q^{-\frac{1}{4}} \,
    \left[
      8 \, q + 24 \, q^2  + 56 \, q^3 + 112 \, q^4
      + \cdots
    \right] \, B_3^{(2)}(z;\tau)
    \\
    & \qquad 
    +
    q^{-\frac{9}{16}} \,
    \left[
      3 \, q + 14 \, q^2 + 28 \, q^3 + 69 \, q^4
      + \cdots
    \right] \, B_3^{(3)}(z;\tau)
    .
  \end{aligned}
\end{gather*}
Polar parts are given by 
\begin{eqnarray}
&&\hskip-1.5cm  \left.\begin{pmatrix}
     \Sigma_{(3,0,0)}^{(1)}(\tau) \\
      \Sigma_{(3,0,0)}^{(2)}(\tau) \\
      \Sigma_{(3,0,0)}^{(3)}(\tau) 
    \end{pmatrix}\right|_{\text{polar}}
    \hskip-7mm =
    \begin{pmatrix}
      116 \, q^{-\frac{1}{16}}
      \\[1mm]
      32 \, q^{-\frac{1}{4}}
      \\[1mm]
      4 \, q^{-\frac{9}{16}}
    \end{pmatrix},
    \hskip3mm 
 \left.  \begin{pmatrix}
      \Sigma_{(2,1,0)}^{(1)}(\tau) \\
      \Sigma_{(2,1,0)}^{(2)}(\tau) \\
      \Sigma_{(2,1,0)}^{(3)}(\tau) 
    \end{pmatrix}\right|_{\text{polar}}
    \hskip-7mm =
    \begin{pmatrix}
      10 \, q^{-\frac{1}{16}}
      \\[1mm]
      q^{-\frac{1}{4}}
      \\[1mm]
      0
    \end{pmatrix},
   \hskip3mm 
  \left.  \begin{pmatrix}
      \Sigma_{(1,1,1)}^{(1)}(\tau) \\
      \Sigma_{(1,1,1)}^{(2)}(\tau) \\
      \Sigma_{(1,1,1)}^{(3)}(\tau) 
    \end{pmatrix}\right|_{\text{polar}}
    \hskip-7mm =
    \begin{pmatrix}
      q^{-\frac{1}{16}}
      \\[1mm]
      0
      \\[1mm]
      0
    \end{pmatrix}.\nonumber \\
    &&
 \end{eqnarray}
We have numerically checked that~\eqref{p_sum_A} with these polar
parts
reproduce above massive coefficients in $Z_{X_3^{(a)}}$.
The theta function in~\eqref{Sigma_and_P} vanishes also in this case.
 
The bases of~\eqref{Eichler_Zagier_base} are written as
% \begin{equation}
%   \begin{pmatrix}
%     Z_{X_3^{(1)}}
%     \\
%     Z_{X_3^{(2)}}
%     \\
%     Z_{X_3^{(3)}}
%   \end{pmatrix}
%   =
%   \begin{pmatrix}
%     \frac{4}{9} & \frac{8}{3} & \frac{8}{9}
%     \\
%     \frac{1}{36} & 0 & - \frac{1}{36}
%     \\
%     \frac{1}{432} & - \frac{1}{144} & \frac{1}{216}
%   \end{pmatrix}
%   \,
%   \begin{pmatrix}
%     \left[ \phi_{0,1} \right]^3
%     \\
%     \left[ \phi_{-2,1} \right]^2 \, \phi_{0,1} \, E_4
%     \\
%     \left[ \phi_{-2,1} \right]^3 \, E_6
%   \end{pmatrix}
% \end{equation}
\begin{equation}
  \label{Jacobi_level3}
  \begin{pmatrix}
    \left[ \phi_{0,1} \right]^3
    \\[1mm]
    \left[ \phi_{-2,1} \right]^2 \, \phi_{0,1} \, E_4
    \\[1mm]
    \left[ \phi_{-2,1} \right]^3 \, E_6
  \end{pmatrix}
  =
  \begin{pmatrix}
    1 & 24 & 96 
    \\
    1 & 0 & -48
    \\
    1 & -12 & 96
  \end{pmatrix}
  \,
  \begin{pmatrix}
    \frac{1}{4} \, Z_{X_3^{(1)}}
    \\[1mm]
    Z_{X_3^{(2)}}
    \\[1mm]
    Z_{X_3^{(3)}}
  \end{pmatrix}
  .
\end{equation}
We note  that the elliptic genus for the Calabi--Yau manifold was
studied in Ref.~\citenum{Gritse99a} where used is
\begin{equation*}
  Z_{X_3^{(3)}}(z;\tau)
  =
  \frac{1}{4} \, \phi_{0,3}(z;\tau) .
\end{equation*}

We denote $X_3(n_2,n_3)$ as
the complex 6-dimensional hyperK\"ahler manifold whose
elliptic genus for $X_3(n_2,n_3)$ is given
by~\eqref{elliptic_genus_general} with $d_3=3$.
We have
\begin{equation}
  \begin{gathered}
    Z_{X_3(n_2,n_3)}(0 ; \tau)
    =
    768 + 48 \, n_2 + 4 \, n_3 ,
    \\[2mm]
    Z_{X_3(n_2,n_3)}\left( \frac{1}{2} ; \tau \right)
    =
    \left( 512 +16  \, n_2 \right) +
    %\left( 512 \, n_2 + 147456 \right) \,q +
    \cdots ,
    \\[2mm]
    Z_{X_3(n_2,n_3)} \left(\frac{1+\tau}{2} ; \tau \right)
    =
    -4 \, q^{-\frac{3}{2}} + \cdots .
  \end{gathered}
\end{equation}
Referring to the generating function~\eqref{DMVV_K3}, we see
$
  K3^{[3]}
  =
  X_3(40, 128)
$,
and we recover the topological invariants as
% \begin{align}
%   Z_{K3^{[3]}}(z; \tau)
%   & =
%   Z_{X_3^{(1)}}
%   +
%   40 \, Z_{X_3^{(2)}}
%   +
%   128 \, Z_{X_3^{(3)}}
%   \nonumber \\
%   & =
%   \frac{50}{27} \, \left[ \phi_{0,1} \right]^3
%   + \frac{16}{9} \, \left[ \phi_{-2,1} \right]^2 \, \phi_{0,1} \, E_4
%   + \frac{10}{27} \, \left[ \phi_{-2,1} \right]^3 \, E_6 ,
% \end{align}
\begin{equation}
  \begin{aligned}
    \chi_{K3^{[3]}} 
    & = 3200,
    &
    \sigma_{K3^{[3]}}
    & = 1152,
    &
    \widehat{A}_{K3^{[3]}}
    & = 4 .
  \end{aligned}
\end{equation}
We note that the character decomposition is given by
\begin{equation*}
  \begin{aligned}
    Z_{K3^{[3]}}(z;\tau)
    & =
    1984 \,  \ch_{3,\frac{3}{4},0}^{\widetilde{R}}(z;\tau)
    - 504 \, \ch_{3,\frac{3}{4},\frac{1}{2}}^{\widetilde{R}}(z;\tau)
    + 64  \,  \ch_{3,\frac{3}{4},1}^{\widetilde{R}}(z;\tau)
    -  4 \,  \ch_{3,\frac{3}{4},\frac{3}{2}}^{\widetilde{R}}(z;\tau)
    \\
    & \qquad 
    +
    q^{-\frac{1}{16}} \,
    \left[
      33880 \, q + 682968 \, q^2 + 9569780 \, q^3
      +
      \cdots
    \right]  \, B_3^{(1)}(z;\tau)
    \\
    & \qquad
    +
    q^{-\frac{1}{4}} \,
    \left[
      -3520 \, q - 375960 \, q^2  - 7364672 \, q^3
      + \cdots
    \right] \, B_3^{(2)}(z;\tau)
    \\
    & \qquad
    + q^{-\frac{9}{16}} \,
    \left[
      -420 \, q +47740 \, q^2 + 1825012 \, q^3
      + \cdots
    \right] \, B_3^{(3)}(z;\tau)
    .
  \end{aligned}
\end{equation*}

\section{Conclusion and Discussion}

In this paper we have studied the general properties of the elliptic
genera of arbitrary  hyperK\"ahler manifolds of complex dimension $2k$.

Using the Rademacher expansion we have shown that the multiplicities
of the (overall) half-BPS states   increase like an exponential and
behaves like 
\begin{equation}
  \exp
  \left(
    2 \, \pi \, \sqrt{k \, n }
  \right)
  \label{entropy_hK}
\end{equation}
for large values of $k$. We would like to identify this phenomenon as
the entropy carried by the  hyperK\"ahler manifolds.

In the standard model of D1-D5 black holes  of string theory
compactified on $K3\times S^1$ with $Q_5$ D5 and $Q_1$ D1 branes, the  effective
theory is a 2-dimensional non-linear $\sigma$-model with the target
space being the symmetric product of
$k=Q_1 \, Q_5$ $K3$ surfaces.
Entropy
of the black hole is given by~\cite{StroVafa96a}
\begin{equation}
  S_{BH}
  =2 \, \pi \,  \sqrt{Q_1 \, Q_5 \, n} ,
  \label{entropy_BH}
\end{equation}
where $n$ is the momentum around $S^1$. We note
that~\eqref{entropy_hK} and~\eqref{entropy_BH}
agree with each other.

Our proposal of the intrinsic entropy for hyperK\"ahler manifolds must
be strengthened by examining similar phenomena in other types of
manifolds: 
we expect that a manifold with a reduced holonomy in general possesses
an intrinsic entropy. In the case of Calabi-Yau  manifolds one uses
the $\mathcal{N}$=2 SCA
and the analysis is more or less similar to the case of hyperK\"ahler
manifolds.
We plan to report on the results of Calabi-Yau manifolds in a subsequent  publication ~\cite{EguchiHikami10a}.
On the other hand, in the case of $G_2$ and spin(7) manifolds the
relevant algebraic structures are not yet known. It is a challenging
problem to develop the representation theory and analyze the elliptic
genera for these manifolds.

\section*{Acknowledgments}
This work is supported in part by Grant-in-Aid from the Ministry of
Education, Culture, Sports, Science and Technology of Japan.

\appendix
\section{Harmonic Maass Form}
The functions $f(\tau)$  is called the harmonic Maass form with weight
$k \in \frac{1}{2}\mathbb{Z}$
on
$\Gamma \in
\left\{
  \Gamma_1(N), \Gamma_0(N)
\right\}
$
if the
followings are fulfilled~\cite{BruinFunke04a,KOno08a};
\begin{itemize}
\item for $\gamma=
  \left(
    \begin{smallmatrix}
      a & b
      \\
      c & d
    \end{smallmatrix}
  \right)\in \Gamma
  $, we have
  \begin{equation}
    f\left(\gamma(\tau)\right)
    =
    \begin{cases}
      (c \, \tau + d)^k \, f(\tau),
      & \text{for $k\in \mathbb{Z}$},
      \\[2mm]
      \displaystyle
      \left(
        \frac{c}{d}
      \right)^{2k} \,
      \epsilon_d^{~-2k} \,
      (c \, \tau + d)^k \,
      f(\tau),
      & \text{for $k\in \mathbb{Z}+\frac{1}{2}$},
    \end{cases}
  \end{equation}
  where $\left( \frac{c}{d} \right)$ is the Legendre symbol, and
  $\epsilon_d$ is defined by
  \begin{equation*}
    \epsilon_d=
    \begin{cases}
      1 & \text{for $d=1 \mod 4$},
      \\
      \I &
      \text{for $d=3 \mod 4$},
    \end{cases}
  \end{equation*}

\item $f(\tau)$ is an
  eigenfunction of the hyperbolic Laplacian~\eqref{Laplacian},
  \begin{equation}
    \Delta_k f(\tau) = 0,
  \end{equation}

\item there exists a polynomial such that
  \begin{equation}
    f(\tau) - \sum_{n \leq 0 } c(n) \, q^n
    =\mathcal{O}
    \left(
      \E^{-\epsilon v}
    \right) ,
  \end{equation}
  for $v=\Im \tau \to \infty$
  and some $\epsilon >0$.
\end{itemize}
%%%%

%%%%%%%%%%%%%%% 
%%%%%%%%%%%%%% %newpage
%\bibliographystyle{physics}
% %\bibliographystyle{amsalpha}
% %\bibliographystyle{JHEP}
% %\bibliographystyle{siam}
% %ibliographystyle{klunum}
\bibliographystyle{alphaKH}
%%%%%%%%%%%%%%%%%%
%\bibliography{_def,gravity,square,math,ba,tba,math5,vm,square2,math4,qalg,math3,math2,poisson,geometry,soliton,cft,knot,tqft,comb,number}

\end{document}